\documentclass[aps,twocolumn,a4paper,showpacs, superscriptaddress]{revtex4}
\usepackage[applemac]{inputenc}
\usepackage{amsmath,amssymb}
\usepackage{graphicx}
\usepackage{float}

\newcommand{\bb}{\bibitem}
\newcommand{\bes}{\begin{subequations}}
\newcommand{\ees}{\end{subequations}}
\def\ben{\begin{eqnarray}}
\def\een{\end{eqnarray}}
\def\be{\begin{equation}}
\def\ee{\end{equation}}
\def\sech{\text{sech}}
\def\sech{\textrm{sech}}

\begin{document}
\title{Scattering of compactlike structures} 
\author{D. Bazeia} 
\affiliation{Departamento de F\'\i sica, Universidade Federal da Para\'iba, 58051-900 Jo\~ao Pessoa PB, Brazil}
\author{T.S. Mendon\c ca}
\affiliation{Instituto de F{\'{\i}}sica, Departamento de F{\'{\i}}sica Te{\'o}rica, Universidade do Estado do Rio de Janeiro, CEP 20550-013. Rio de Janeiro, RJ, Brazil.}
\author{R. Menezes}
\affiliation{Departamento de F\'\i sica, Universidade Federal da Para\'iba, 58051-900 Jo\~ao Pessoa PB, Brazil}
\affiliation{Departamento de Ci\^encias, Universidade Federal da Para\'iba, 58051-900 Jo\~ao Pessoa PB, Brazil}
\author{H.P. de Oliveira}
\affiliation{Instituto de F{\'{\i}}sica, Departamento de F{\'{\i}}sica Te{\'o}rica, Universidade do Estado do Rio de Janeiro, CEP 20550-013. Rio de Janeiro, RJ, Brazil.}

\begin{abstract}
We investigate the collision of a new class of topological defects that tends to become compact as a control parameter increases to larger and larger values These new compactlike defects have, in general, more than one internal discrete mode depending on the value of the control parameter and, as usual, there is a critical velocity above which the defects escape after the collision. We noticed that below the critical velocity there are the windows of escape presenting fractal structure. An interesting novelty is the appearance of metastable structures with the formation of compactlike defects, maintaining a fixed distance from each other.  Another new feature is the formation of boosted localized distributions of the scalar field which we called moving oscillons. These oscillons carry away almost all scalar field energy producing a complete disruption of the compactlike defects. The pattern of the moving oscillons depends on the control parameter, and becomes more complex as we increase its value. We conjecture that the new effects may be connected with the presence of more than one vibrational mode in the spectrum of the stability potential of the model under investigation.
\end{abstract}
\maketitle

\section{Introduction}

Topological defects are of current interest in high energy physics, where they can be used to describe phase transitions in the early universe or map interfaces separating distinct regions in space, among several other possibilities \cite{r1,VS,MS,Vacha,Shnir}. Some interesting illustrations of the use of topological defect appeared before, for instance, in Refs. \cite{AV,DM,DS,AI,TR,DMR} and in references therein.

In this work we study topological defects of the kink type, which appear in relativistic models described by a real scalar field in $(1+1)$ spacetime dimensions. We focus attention on models that support kinks with the compact profile, but we recall that kinks with the standard profile appeared before in the $\phi^4$ model. Although compact structures first appeared in the presence of nonlinearity and nonlinear dispersion, acquiring spatial profile with compact support \cite{RH}, here we will follow another route \cite{b}, in which one shows how to get topological structures that may acquire the compact profile in a scalar field model with standard kinematics. We work with natural units $(\hbar=c=1)$ and use dimensionless field and space and time coordinates.

As one knows, compact structures behave trivially outside a compact region, so they are different from the standard kinklike configurations and their collisions may bring new effects. This is the main aim of the current work, and we will compare the results with the scattering or collisions of kinks that have been investigated in Refs.~\cite{campbell,sca1,sca2,tiago,sca4,sca5,sca6,sca7,sca8,bagomes} and in references therein. 
In \cite{campbell}, in particular, the authors have suggested that the two-bounce windows that appeared in the collisions of kinks in the
$\phi^4$ model are due to the existence of the translational and vibrational modes. Motivated by this, in the present work we consider a model that may support several vibrational states, so the scattering of kinks may have a richer structure. In fact, we study the scattering of kinks in a model controlled by an integer $n=1,2,3, \cdots$, which leads to the compact limit for $n\to\infty$, where the kinks shrink to a compact interval of the real line. We do this to examine how the scattering behave as we increase the parameter that controls the scalar field theory, since this will contribute to increase the number of vibrational states in the system. We organise the work as follows: in Sec.~\ref{sec-cla} we briefly review the model introduced in \cite{b}, which is capable of describing compact solutions in a model with a single real scalar field in $(1+1)$ spacetime dimensions. In Sec.~\ref{coll} we consider the collisions of the compactlike configurations presented in Sec.~\ref{sec-cla} and collect the mains results of the numerical investigations. We conclude the work in Sec.~\ref{sec-com}.

\section{The model}\label{sec-cla}

In order to investigate the problem, let us write the Lagrange density for a scalar field with standard kinematics. It has the form
\be\label{model}
{\cal L}=\frac12{\dot\phi}^2-\frac12\phi^{\prime2}-V(\phi),
\ee
where we are using $\dot\phi=\partial\phi/\partial t$, $\phi^\prime=\partial\phi/\partial x$, the metric is $(+,-)$ and $\phi$ is a real scalar field. $V(\phi)$ is the potential and the equation of motion is given by
\be\label{eqMotion1}
\ddot\phi-\phi^{\prime\prime} + \frac{dV}{d\phi}=0.
\ee
In this work we will consider a non negative potential $V(\phi)$ that supports two minima, at $\bar\phi_\pm=\pm1$, and a single topological sector. It supports the kinklike solution $\phi(x)$, which connects the two minima: $\phi(x\to-\infty)=-1$ and $\phi(x\to\infty)=1$. The kinklike solution obeys the equation
\be
\phi^{\prime\prime}=\frac{dV}{d\phi},
\ee
with $\phi(x\to -\infty)\to -1$ and $\phi(x\to\infty)\to 1$, with vanishing derivatives.
It has energy density $\rho(x)$ given by 
\be
\rho(x)\!=\!\frac12 \phi^{\prime2}+V(\phi(x))=\phi^{\prime2}=2V(\phi(x)).
\ee

We study linear stability, to see how the solution behaves under the presence of small fluctuations. We add small fluctuations around the static solution $\phi(x)$, writing $\phi(x,t)=\phi(x)+\eta(x)\cos(\omega t)$. We use this into the equation of motion and expand it up to first-order in $\eta$ to get the Schroedinger-like equation 
\be \label{stab}
\left(-\frac{d^2}{dx^2}+U(x) \right)\eta=\omega^2 \eta,
\ee
with 
\be\label{qm}
U(x)=\left.\frac{d^2V}{d\phi^2}\right|_{\phi=\phi(x)}.
\ee
This is the potential that appears in the study of stability, so we call it the stability potential. 
We see from the above Eq.~\eqref{qm} that $U(x)$ goes asymptotically to $m^2$, which is the (squared) classical mass of the field at the minima $\pm1$.

An important model which falls within the class of models that we are interested in is the $\phi^4$ model, with spontaneous symmetry breaking. The potential is the prototype of the Higgs field can be written as
\be\label{p4}
V(\phi)=\frac12(1-\phi^2)^2.
\ee
We focus on this model, which has the minima
$\pm 1$ and a maximum at the origin, such that $V(0)=1/2$. We also have the kink solution given explicitly by
\be\label{k}
\phi(x)= \tanh( x),
\ee
where we are taking the center of the kink at the origin, for simplicity. The energy density has the form
\be\label{ed}
\rho=\sech^4(x),
\ee
and the energy gives $E=4/3$. In this case, the stability potential has the form
\be
U(x)=4-6 \,{\rm sech}^2(x).
\ee
It is the modifield Poeschl-Teller potential and is reflectionless and has two bound states, the zero mode, with $\omega_0=0$, and the excited state, with $\omega^2_1=3$.

Let us now consider another model, given by
\be\label{ln}
{\cal L}_n=\frac12 \partial_\mu\phi\partial^\mu\phi-V_n(\phi),
\ee
where the potential has the form
\be\label{potmodel2}
V_n(\phi)=\frac12 \left(1-\phi^{2n} \right)^2,
\ee
with $n$ being an integer, $n\geq1$.  For $n=1$, we get back to the $\phi^4$ model. This model is different from the previous one, and the parameter $n$ is now a natural number that controls the model.

\begin{figure}[t]
\includegraphics[width=4.2cm]{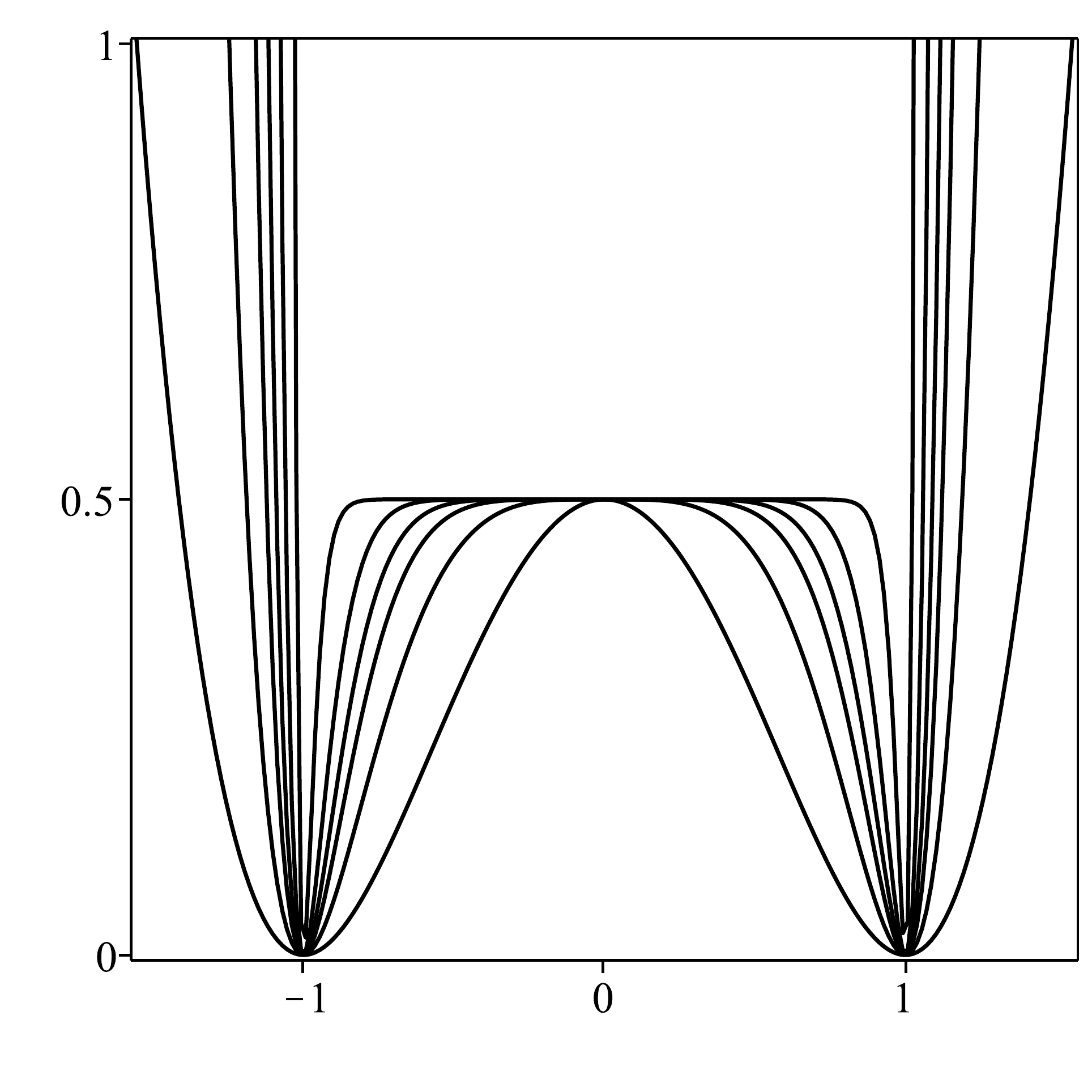}
\includegraphics[width=4.3cm]{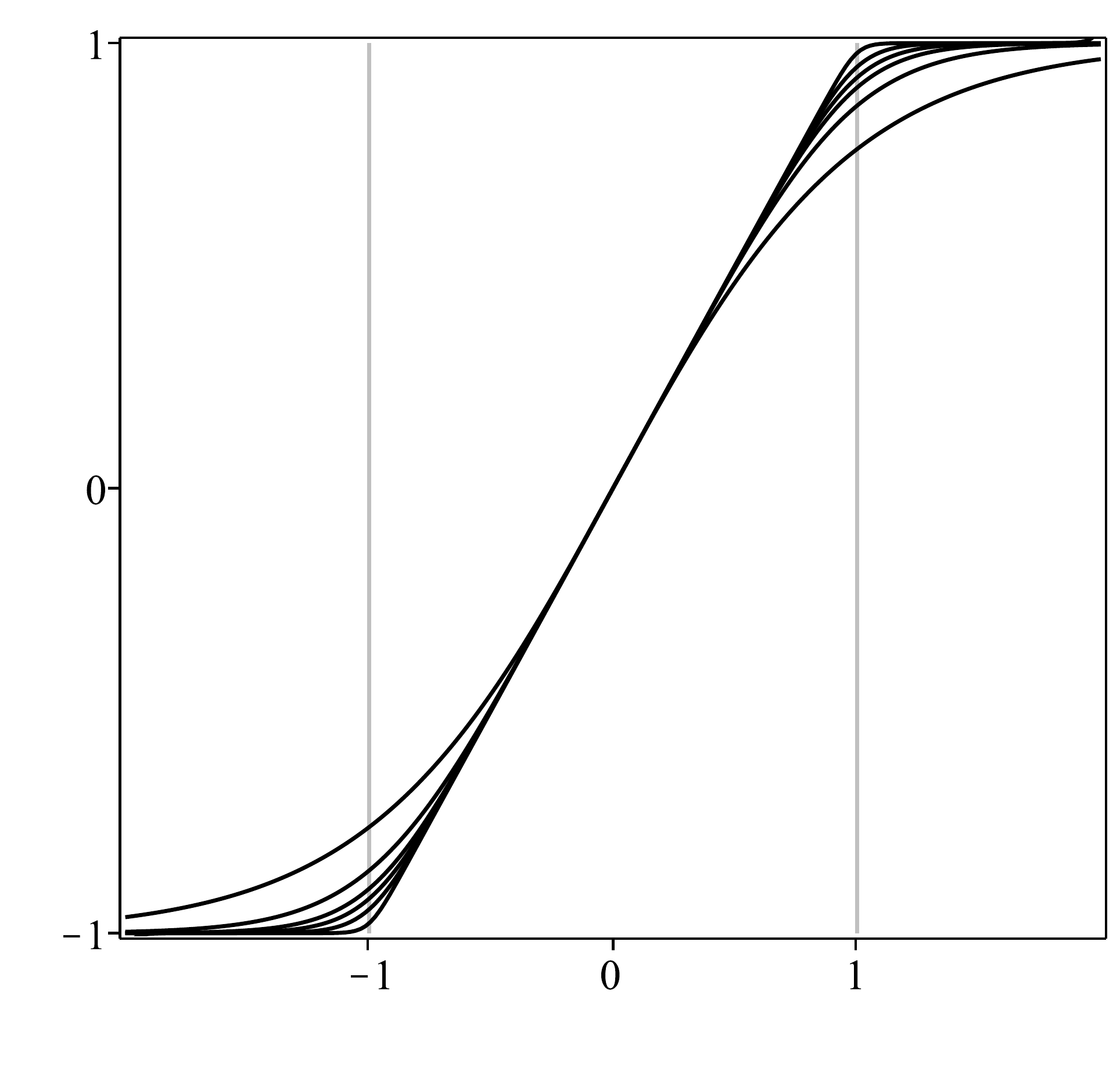}
\includegraphics[width=4.2cm]{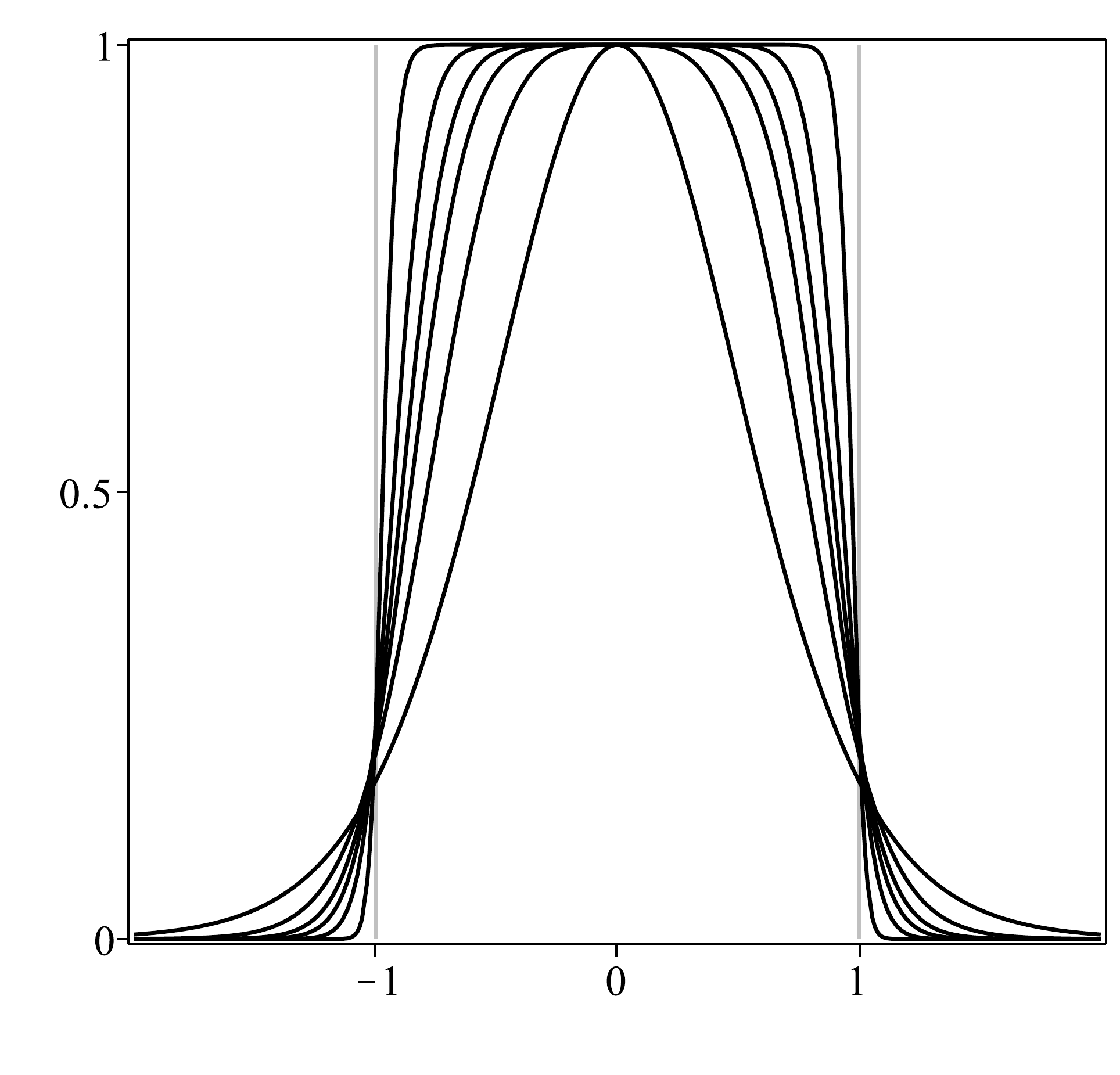}
\includegraphics[width=4.2cm]{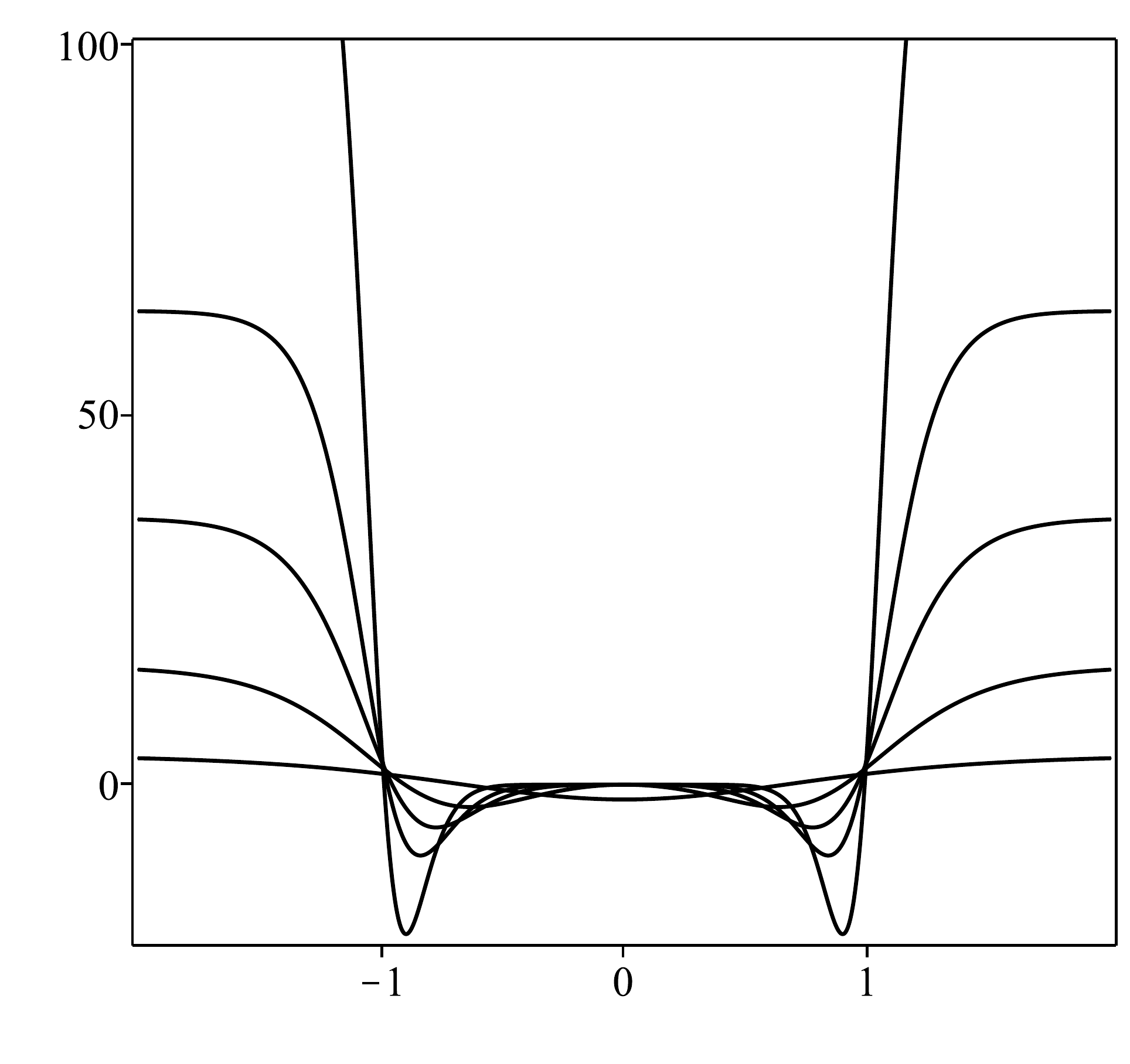}
\caption{The potential \eqref{potmodel2} (top left), its corresponding kinklike solution (top right), energy density (bottom left) and stability potential (bottom right), depicted for $n=1$ and for several other higher values of $n$.}
\label{fig1}
\end{figure}

The equation of motion for this model is
\be\label{eomn}
\phi^{\prime\prime}=-2n\phi^{2n-1}(1-\phi^{2n}).
\ee
The second derivative of the potential is
\be
\frac{d^2V_n}{d\phi^2}=4n^2\phi^{4n-2} - 2n(2n-1)\phi^{2n-2}(1-\phi^{2n}).
\ee
This potential is non-negative and also has the two minima $\phi_\pm=\pm1$, with $V_n(\pm1)=0$, for any $n$. Also, the mass is such that $m_n^2=4n^2$, which increases to very large values for very large values of $n$. 

The equation of motion \eqref{eomn} can be solved numerically, and in Fig.~\ref{fig1} we depict the potential (top left), kinklike solution (top right), energy density (bottom left) and stability potential (bottom right) for several values of $n$. We see from the stability potential that it changes with increasing $n$, becoming a well with increasing walls that gets more and more bound states, as $n$ increases to larger and larger values. One also notes that, as $n$ increases to larger and larger values, the kinklike solution approaches the compact limit; in this case, the solution tends to become a straight line that goes from $-1$ to $1$ in the compact interval $[-1,1]$ and the energy density shrinks to be nested inside the same interval.   

\begin{figure}[h]
	\includegraphics[width=7cm,height=4.5cm]{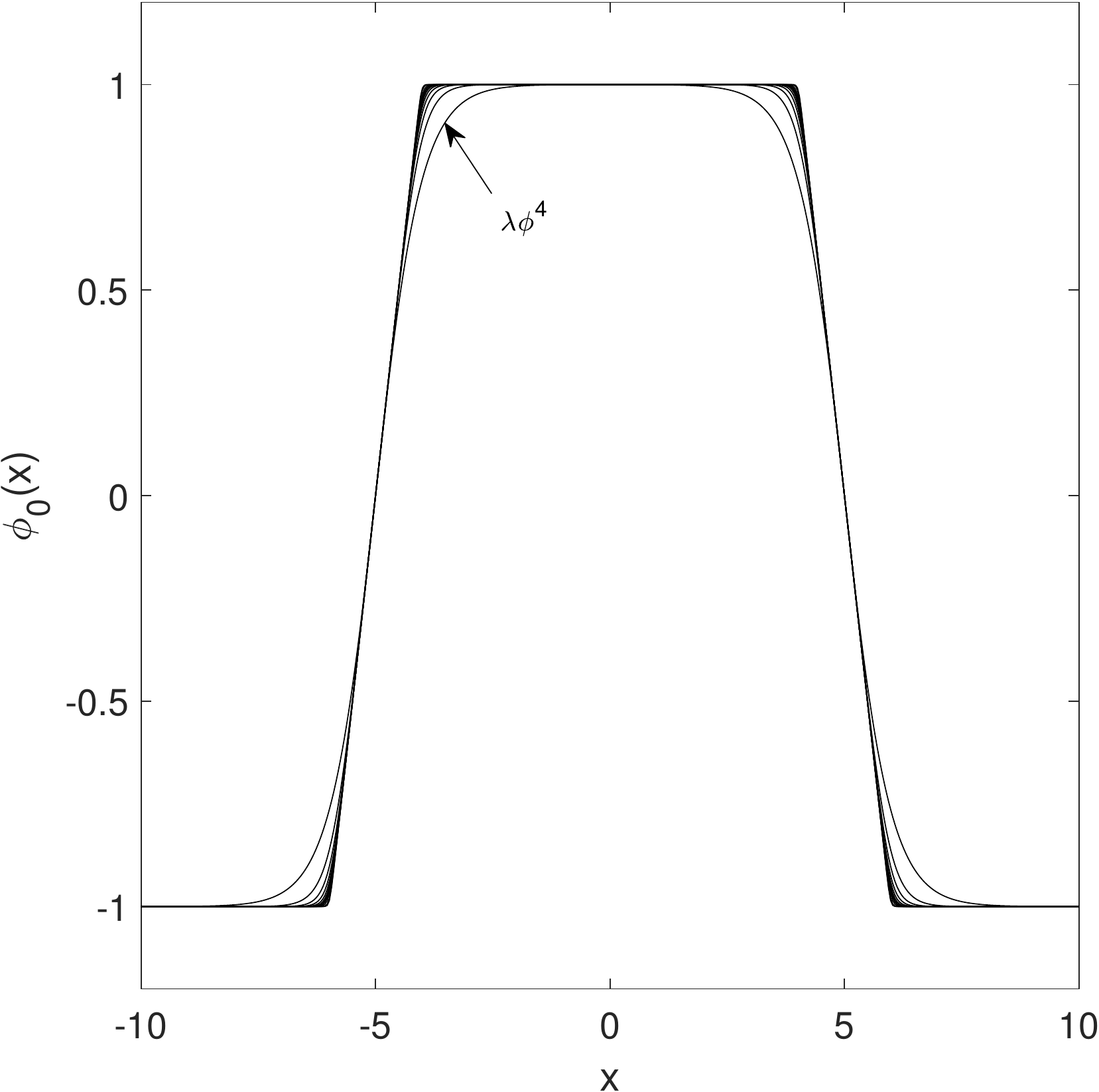}
	\includegraphics[width=7cm,height=4.5cm]{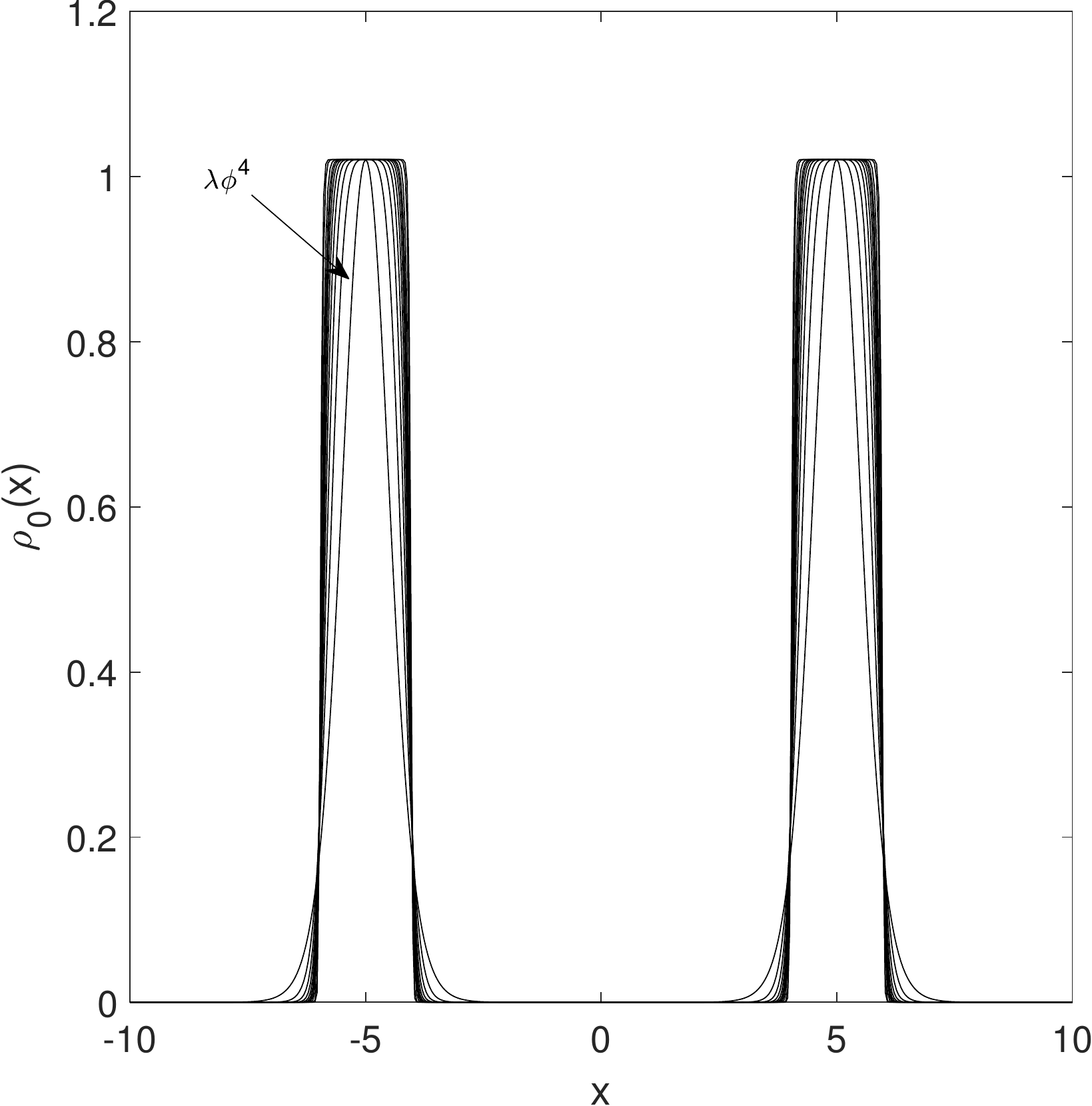}
	\caption{(Top panel) The profile of the initial data (15) for successive values of $n$ varying from $n=1$ corresponding to the $\phi^4$ model, up to $n=20$. (Bottom panel) The initial configuration of the energy density corresponding to the configuration in the top panel.}
\end{figure}

\section{Scattering}\label{coll}

Let us now investigate the scattering of the kink-like structures or the new topological defects that appear in the model described by Eq. (12), for some specific values of $n$. The case $n = 1$ reproduces the $\phi^4$ model, and we use it as a standard situation. We refer to the topological defects for $n=1$ as kinks, and we distinguish them from the new kinklike structures that we call compactlike defects characterised by $n>1$. We will consider several values of $n>1$ to investigate how the usual scattering of the $\phi^4$ model changes when one increases $n$, remembering that the case $n\rightarrow \infty$ leads to the compact kinks. 

We consider an initial configuration that represents a pair of the compactlike defects moving towards each other with velocity $u$. At $t=0$, we have

\begin{equation}\label{init}
\phi_0(0,x) = -1 + \phi_C(0,x+x_0) - \phi_C(0,x-x_0),
\end{equation}

\noindent where $2 x_0$ is the initial distance between the two kinklike centers, located at $x=- x_0$ and $x=x_0$, respectively. Also, we need to include the initial condition $v_0(0,x)=\left(\partial \phi/\partial t \right)_{t=0}$. In this case, it is necessary to extend the static compact-like solution to include the boost factor \cite{r1,Vacha}. However, it is not possible to obtain an explicit exact solution for a boosted compactlike solution, $\phi_C(t,x)$ ($n>1$),  unless for the case $n=1$, which describe the standard kinks. We circumvent this problem by obtaining the implicit solution after integrating Eq. (3) with the introduction of a boost factor, which results in

\begin{equation}
\int\,\frac{d \phi}{1-\phi^{2n}} = \frac{x-x_0-u t}{\sqrt{1-u^2}}.
\end{equation}

\noindent After selecting $n=2,3,4,$ etc, we integrate the LHS of the above equation to obtain the corresponding implicit solution of a compactlike defect located initially at $x=x_0$ and moving with speed $u$ in the positive $x$ direction if $u>0$. For the sake of illustration,  we show in Fig. 2 the initial profiles of the scalar field and the energy density corresponding to the initial data \eqref{init} with $x_0=5$ and $u=0.15$ for $n=1,2,..,8$. 

We integrate the equation of motion \eqref{eqMotion1} numerically with the spectral code introduced in Ref. \cite{tiago}. Briefly, we approximate the scalar field as 

\begin{equation}
\phi(t,x)=\sum_{j=0}^{N}\,a_j(t) \psi_k(x)
\end{equation}

\noindent where $a_j(t)$ are the unknown coefficients associated to the basis functions $\psi_j(x)$ and $N$ is the truncation order whose value coincides with the number of collocation points \cite{tiago}. The equation of motion is then reduced to a set of $N+1$ coupled ordinary differential equations for the modes $a_j(t)$. We improved the code resolution by setting $N \geq 1000$ in such a way that the relative error in the total energy $\delta E(t) = |E_{\mathrm{initial}}-E(t)|/E_{\mathrm{initial}} \times 100$ is not superior to $10^{-4}\%$, where the total scalar field energy is given by 

\begin{equation}
E(t) = \int_{-\infty}^\infty\!\!\!\!dx\left(\frac{1}{2}\left(\frac{\partial \phi}{\partial t}\right)^2+\frac{1}{2}\left(\frac{\partial \phi}{\partial x}\right)^2+V(\phi)\right).
\end{equation}

\subsection*{Collision of kinks} 

The collision of kinks of the $\phi^4$ model ($n=1$) results in two primary outcomes depending on the initial impact velocity. For velocities higher than the critical value, $u_{\mathrm{crit}}$, the kinks interact briefly and emerge moving apart from each other. On the other hand, for $u <  u_{\mathrm{crit}}$, in general, there is a formation of a bound state constituted by an oscillating structure at the origin. The interaction between the kinks is almost elastic with a small loss of the scalar field.

\begin{figure}[hb]
\includegraphics[width=6.7cm,height=4.5cm]{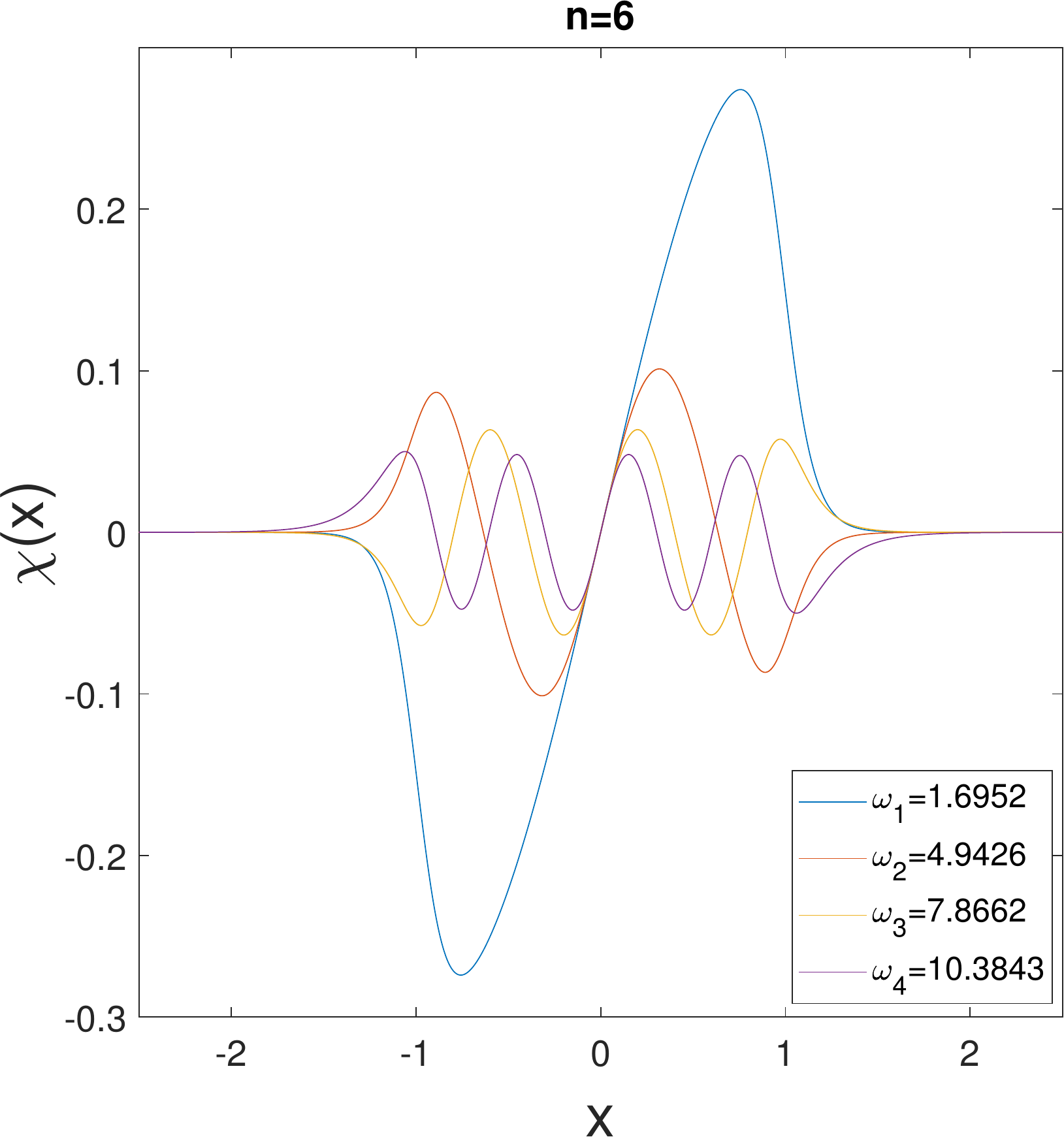}
\includegraphics[width=6.7cm,height=4.5cm]{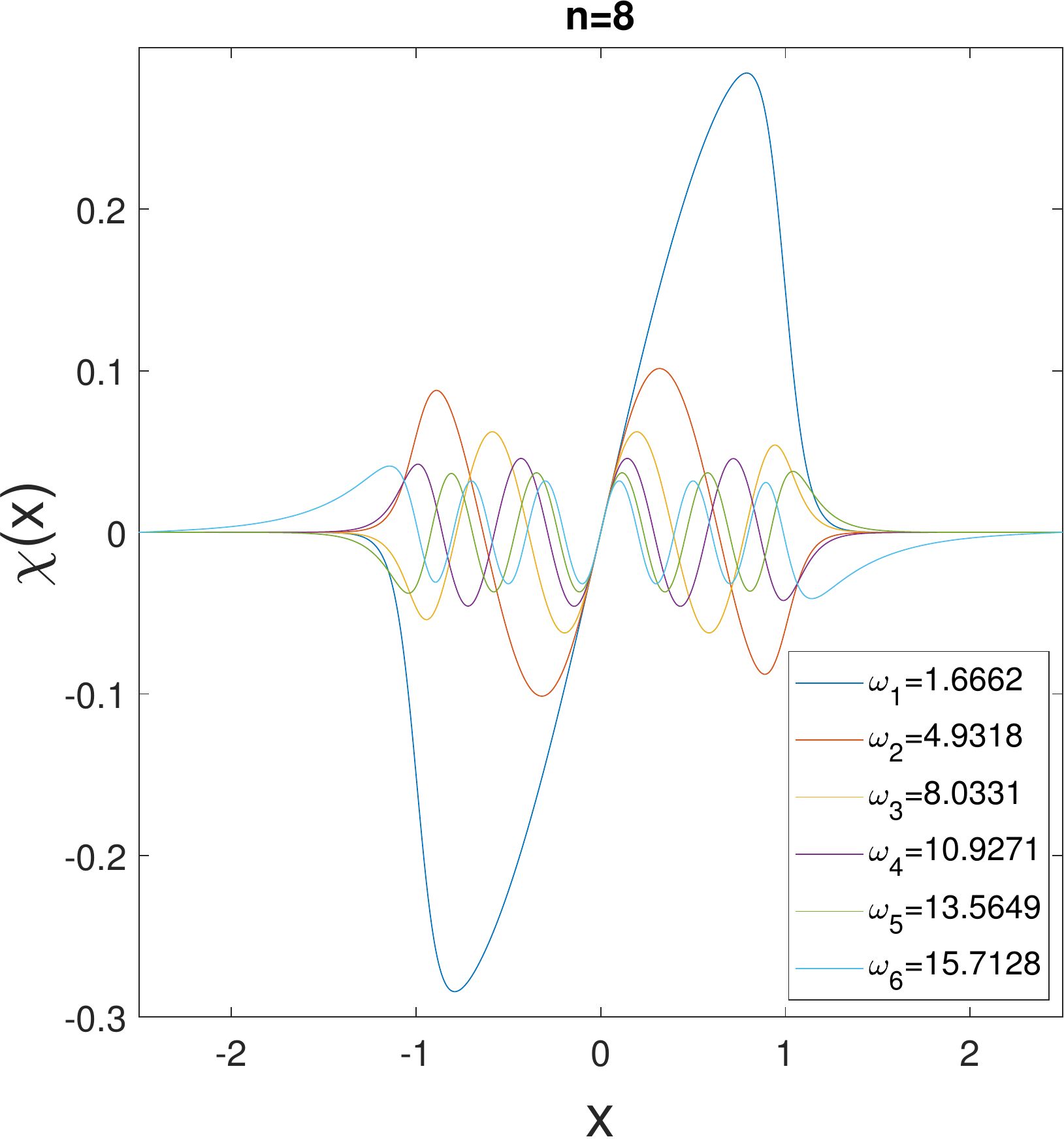}
\caption{The eigenfunctions $\chi(x)$ associated to the internal modes of the compactlike structures with $n=6$ and $n=8$. In the inset, we display the corresponding eigenfrequencies.}
\end{figure}
\begin{figure*}
	\includegraphics[scale=0.35]{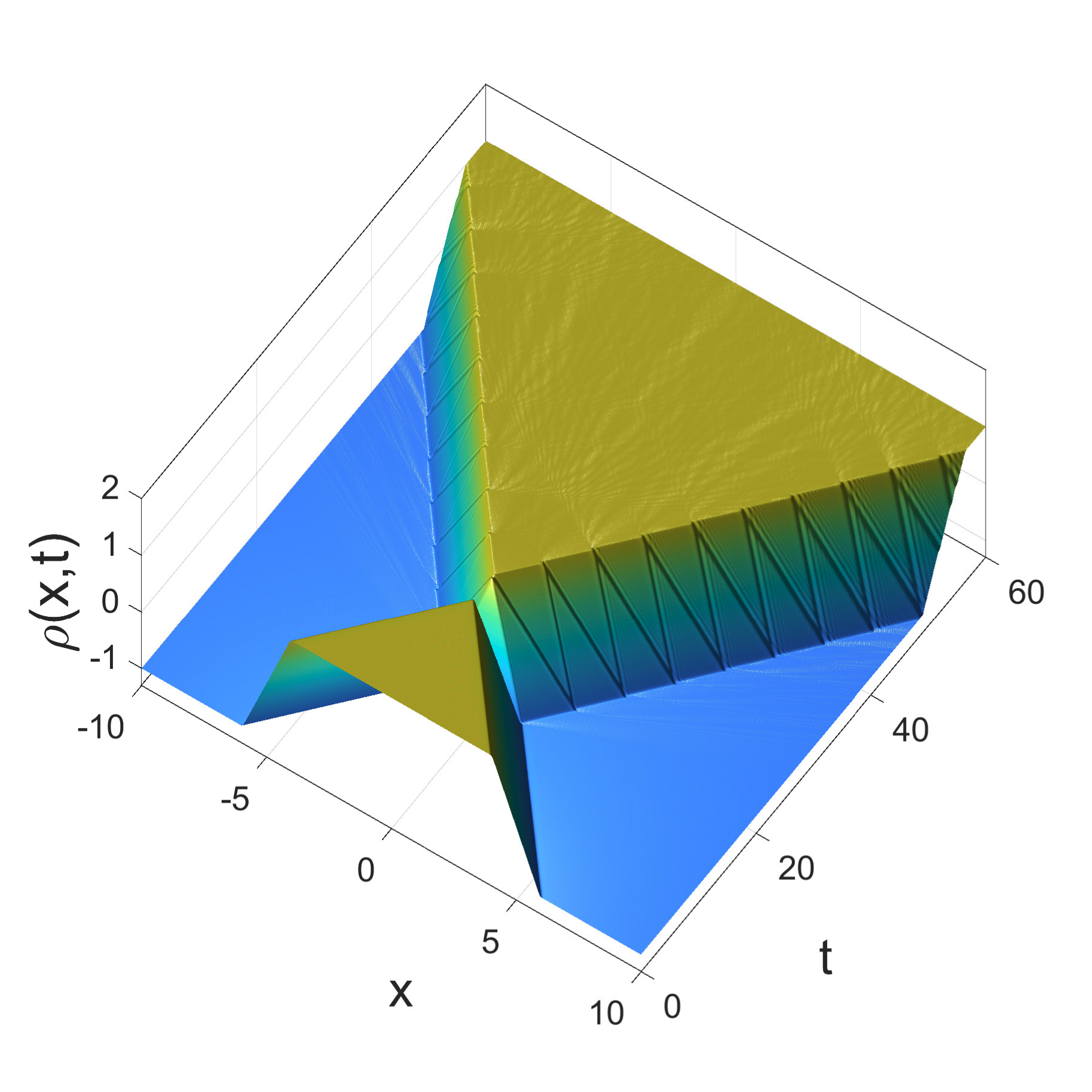}\includegraphics[scale=0.35]{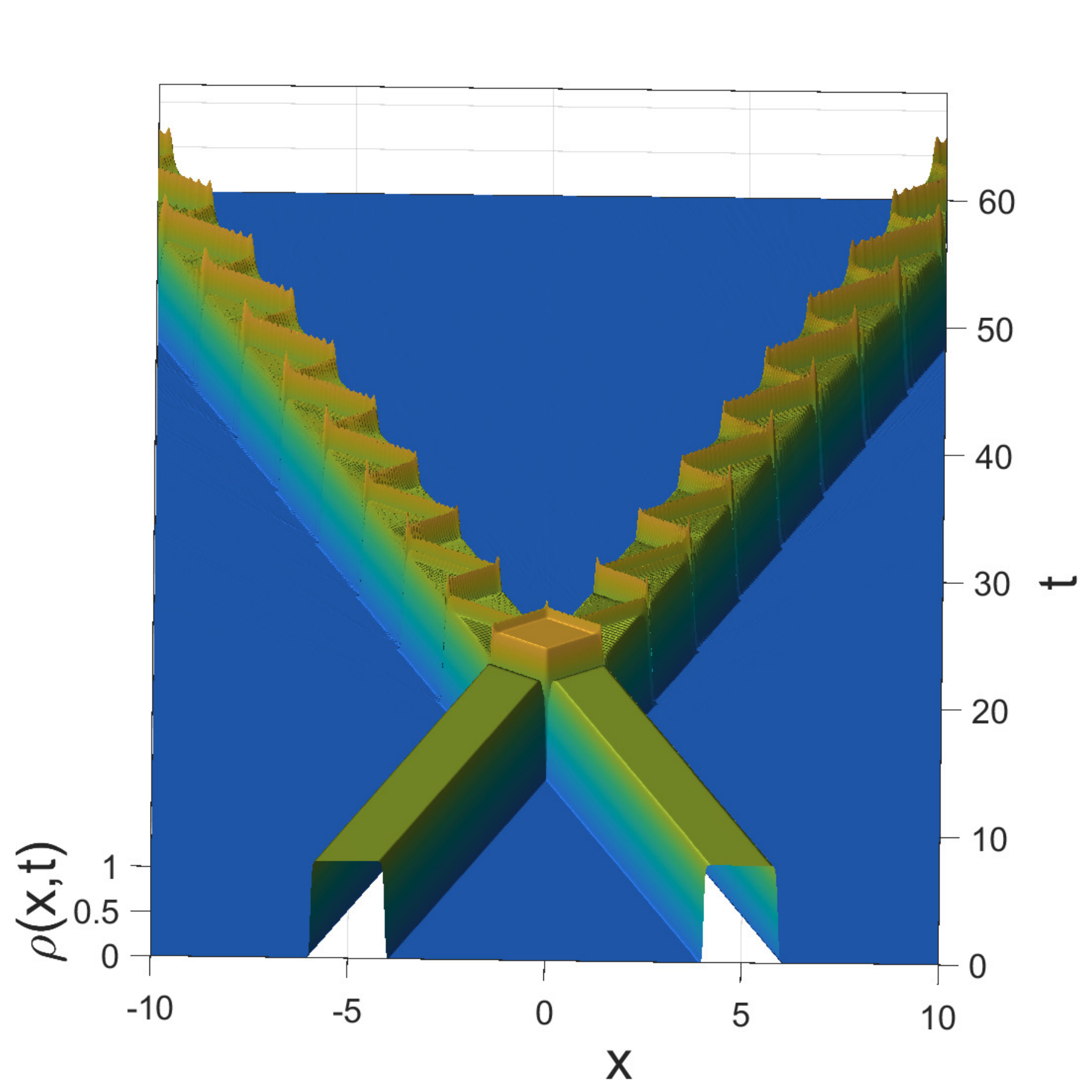}
	\caption{3D plots of the scalar field and the energy density of the escape of the compactlike kinks for $u>u_{\mathrm{crit}}$. Here $n=20$ and $u=0.278577$. }
\end{figure*}

As shown in several works \cite{makhankov,moshir,sugiyama,campbell,aninos,goodman}, the remarkable nonlinear aspect of the interaction of kinks manifests with the presence of the windows of escape or the resonance windows. It means that for $u <  u_{\mathrm{crit}}$ there exist some narrow intervals in which the kinks emerge and escape after performing one or more bounces. The best explanation for the appearance of the windows of escape is the reversible exchange of energy between the vibrational and translational modes of the kink and antikink. When the kinks collide, they separate, but part of their kinetic energy is transferred to the vibrational mode supported by the stability potential. Then, the kinks emerge and due to their mutual attraction return to collide again since $u < u_{\mathrm{crit}}$. This process can be repeated several times, giving rise to the bounces. However, the energy stored in the oscillatory modes can be transferred to the translational modes resulting in the escape of the kink and antikink. This mechanism is not particular for the $\phi^4$ model, but it is extended to other similar models.  Furthermore, the edges of the windows of escape are not smooth but have a fractal structure.

\subsection*{Internal modes of the compactlike structures} 

In general, the internal modes of a topological defect may play an essential role when the defects interact. For this reason, it is of interest to identify the internal modes of the stability potential of the present model for $n$ increasing from unity to higher values. 

For the sake of completeness, we obtained the internal modes of a compactlike defect following the standard procedure \cite{Vacha,sugiyama}. We first perturb the compactlike structure as $\phi(x,t)=\phi(x)+\chi_k(x)\cos(\omega_k t)$. After obtaining the linearised Sch\"orindger-like equation, we solve it for the eigenfunctions $\chi_k(x) $ and the corresponding eigenfrequencies $\omega_k$. For the case of $n=1$, there is one discrete mode identified as the vibrational mode with eigenfrequency $\omega=\sqrt{3}$, besides the translational or zero mode characterised by $\omega=0$. Considering now the compactlike structures $(n>1, \rm {integer})$, we found that there exist, in general, more than one discrete mode besides the zero mode. Moreover, the continuous radiative spectrum is characterised by $\omega_k \geq 2 k$ for all $k=1,2,3,$ etc, with $k$ controlled by $n$. We present the plots of some eigenfunctions for the cases $n=6$ and $n=8$ in Fig. 3. As a consequence, we expect new features arising from the interaction of the compactlike structures when the impact velocities are smaller than the critical velocity due to a possible complex transfer of energy among the distinct internal modes. In Table 1 we present the single precision discrete eigenfrequencies for some values of $n$. 

\begin{table}
	\centering
	\begin{tabular}{|c|c|}
		\hline
		$\,n\,$ & $\omega_k$  \\
		\hline
		\hline
		$1$ & $\sqrt{3} \approx 1.7321$ \\
		\hline 
		$2$ & $1.8421$,\, $3.8366$ \\
		\hline
		$3$ & $1.7896$,\, $4.6036$ \\
		\hline
		$4$ & $1.7472$,\, $4.8465$,\, $7.1687$ \\
		\hline
		$5$ & $1.7170$,\, $4.9231$,\, $7.644$,\, $9.6901$\\
		\hline
		$6$ &\, $1.6952$,\,  $4.9426$,\, $7.8662$,\, $10.3842$\,\\
		\hline
	\end{tabular}
	\caption{Values of the discrete eigenfrequencies corresponding to $n=1,..,6$.}
\end{table}

\subsection*{Critical velocity and escape of compactlike kinks} 

In the case of the collision of compactlike defects, there also exists a critical velocity defined for each $n>1$. We found that the critical velocity increases slightly with the parameter $n$, for instance, for the sake of illustration, we obtain the sequence of some critical values as $u_{\mathrm{crit}} \approx 0.25104\;(n=2), \,0.26052 \;(n=3), \,0.26357 \;(n=4), \,0.26573\;(n=5),\,0.26733\;(n=6)$.

For the impact velocity higher than the critical speed, the compactlike structures reflect and move apart from each other similarly to the reflection and escape of kinks. We illustrate this situation in Fig. 4 with the 3d plots of the scalar field and energy density for the collision of compactlike kinks with $n=20$ and initial velocity of $u=0.278577$. We notice that the receding compactlike defects oscillate as the result of the excitement of their internal vibrational modes. It is also clear that the presence of compact structures before and after the collision.

	\begin{figure*}[ht]
		\includegraphics[width=9cm,height=3.5cm]{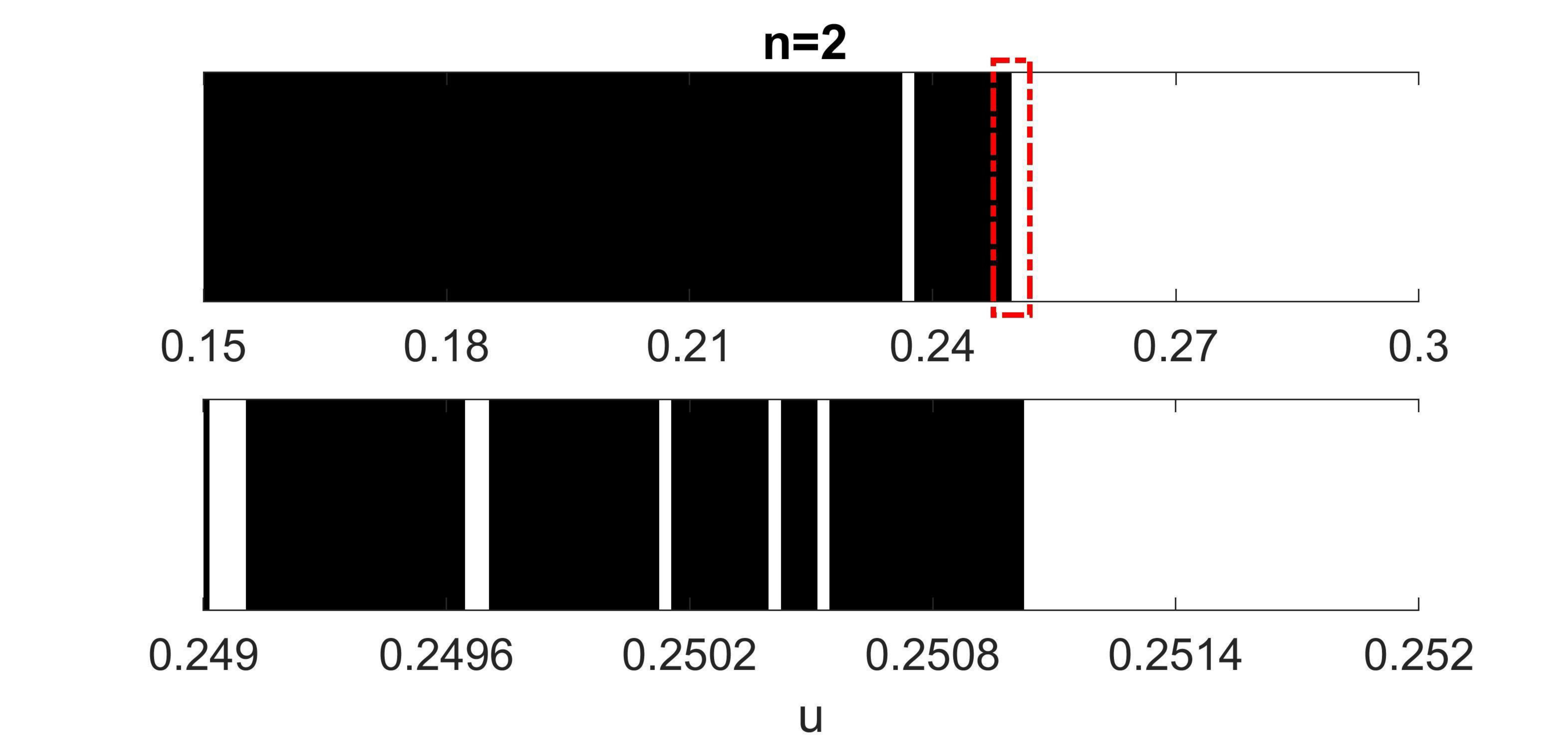}\includegraphics[width=9cm,height=3.5cm]{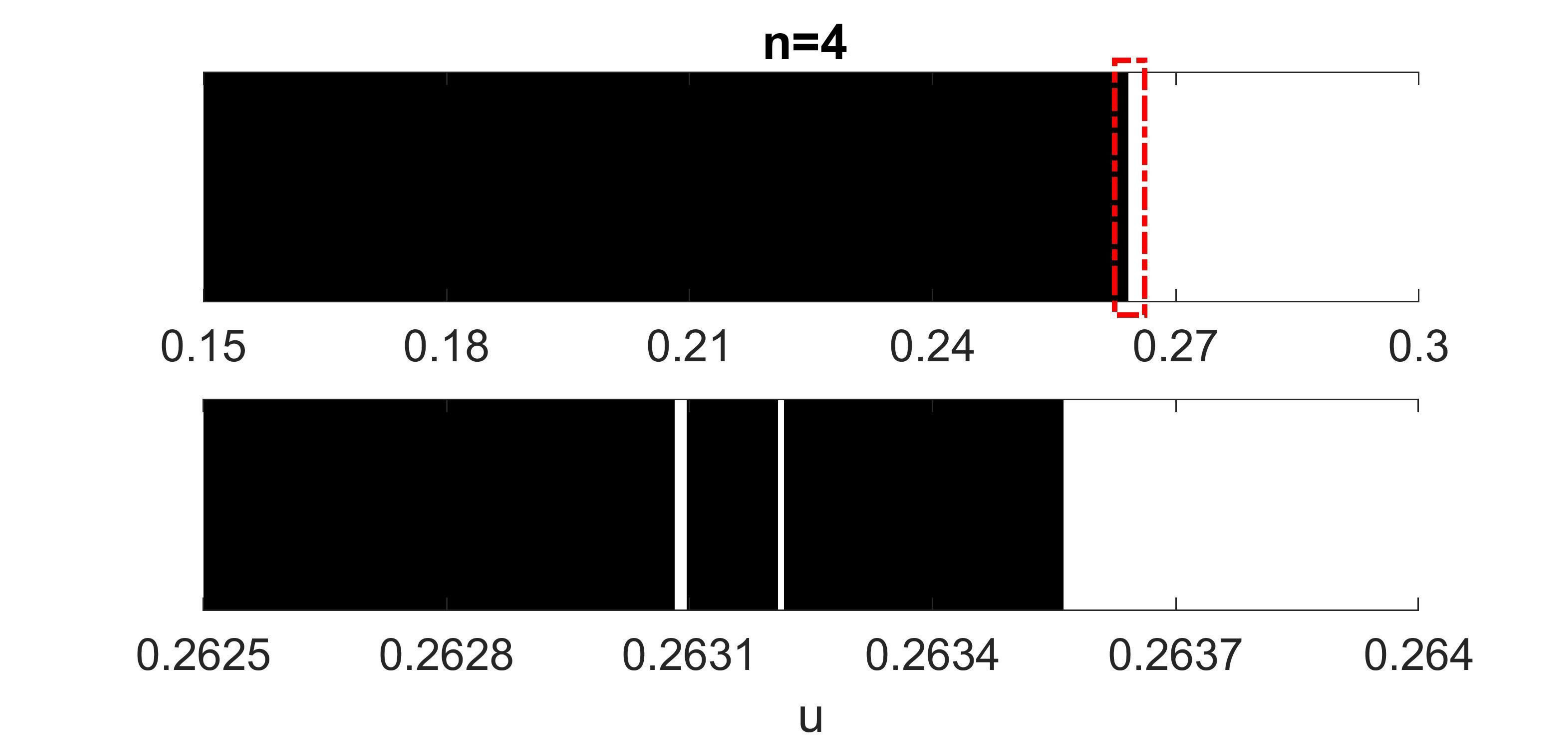}\\
		\includegraphics[width=9cm,height=3.5cm]{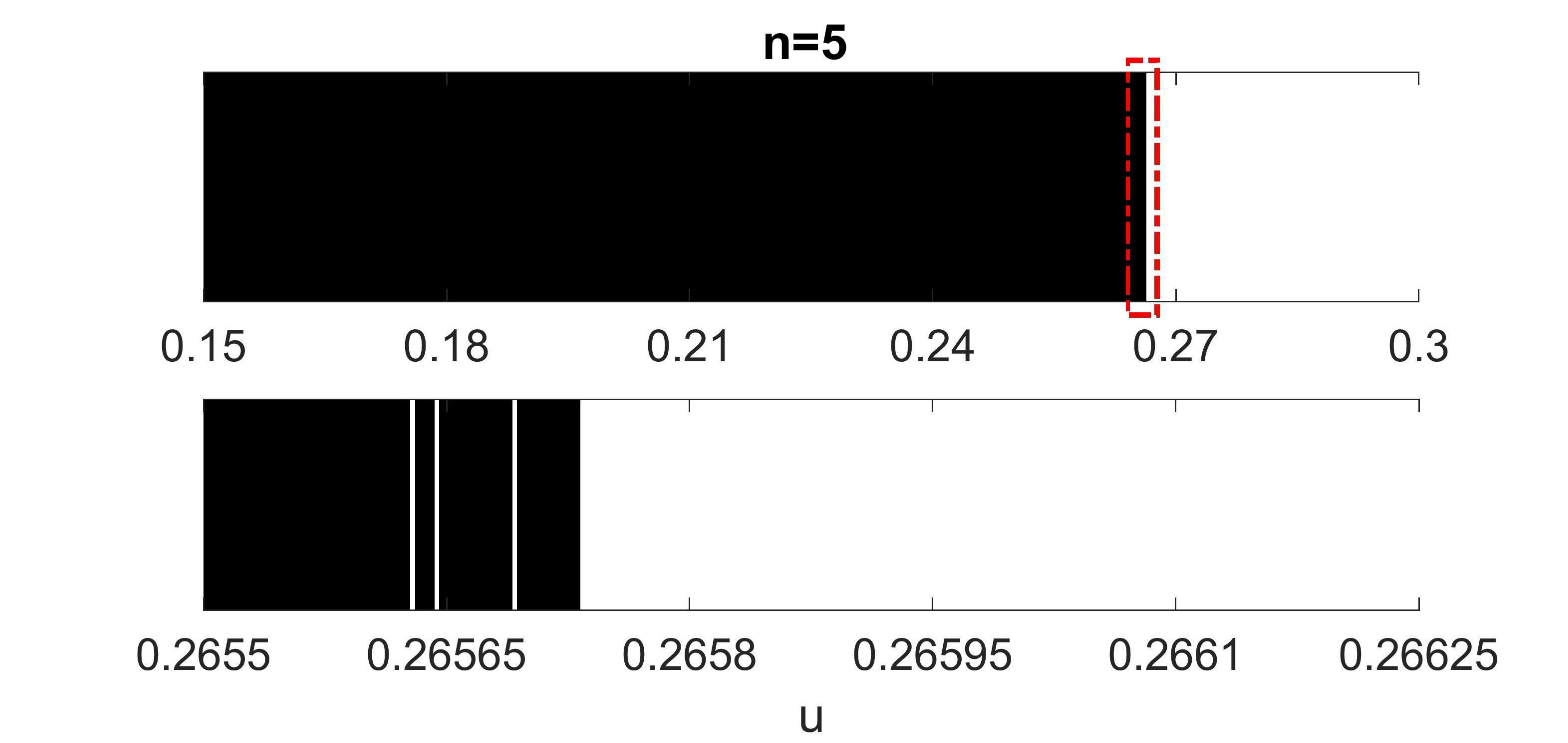}\includegraphics[width=9cm,height=3.5cm]{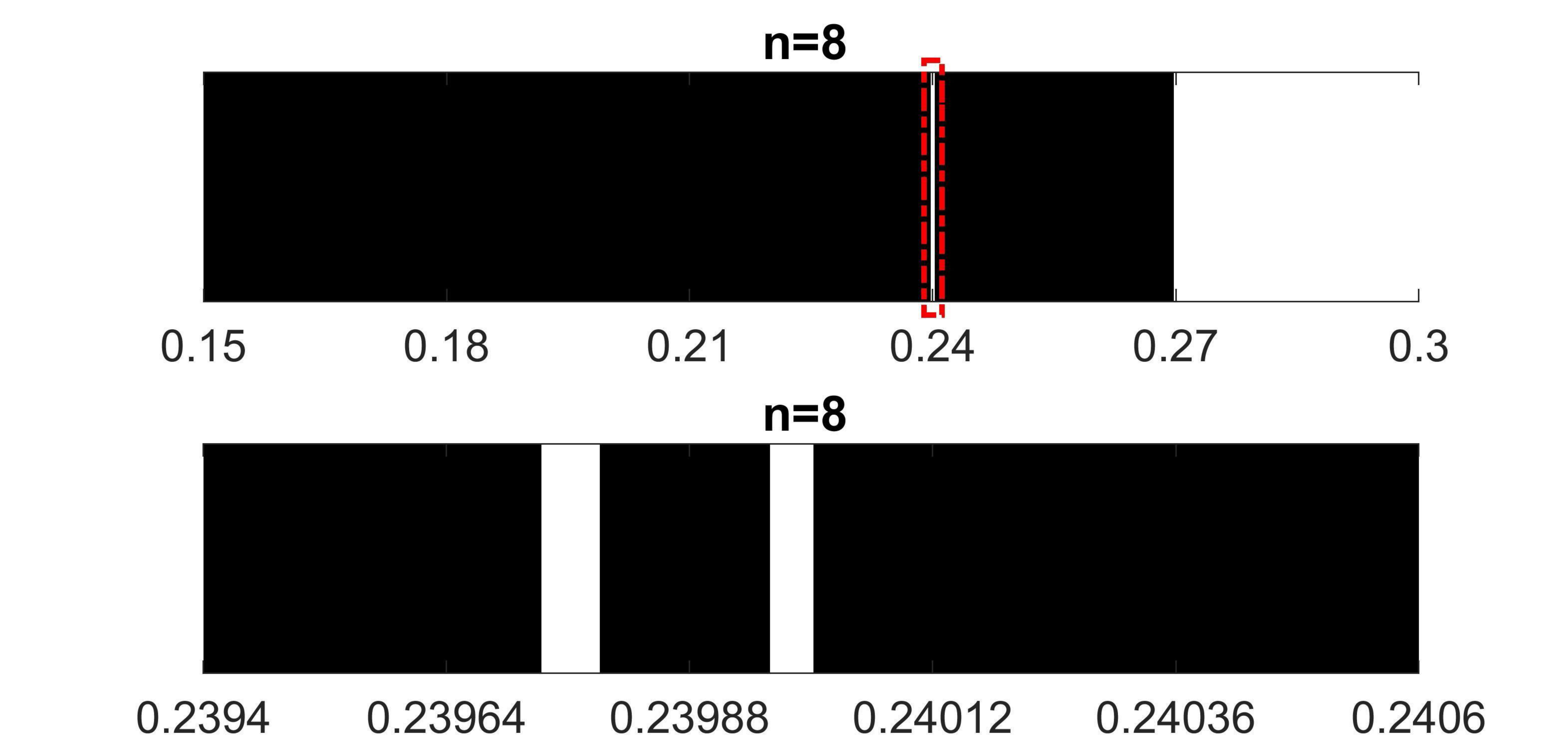}
		\caption{illustrations of the fractal structure by zooming in some of the boundaries of the windows of escape for the collision of compactlike kinks with distinct values of $n$.}
	\end{figure*}

\subsection*{Fractal structure}

The numerical experiments indicated the existence of fractal structures in the edges of the windows of escape for any $n > 1$. We show in Fig. 5 the illustration of the fractal structure by zooming in the boundaries of some windows of escape, where white indicates the escape of the compactlike defects. Like the collision of kinks, it is a robust nonlinear aspect of the interaction of compact-like defects. However, the interaction of compactlike defects exhibits striking differences with respect to the kinks ($n=1$) as we are going to describe.

In connection with the fractal structure, the presence of the windows of $2,3,4,..,k$-bounces or resonance windows are ubiquitous. From the collision of kinks, we know that the bounces are a consequence of an intricate exchange of energy stored into the vibrational and translational internal modes along with a resonance mechanism \cite{campbell,aninos,goodman}. In the collision of compactlike defects, the transfer of energy among the distinct vibrational models produces a rich structure of resonance windows. For the sake of illustration, we show in Fig. 6 examples of 2-, 4- and 8-bounce solutions or resonances for the compactons with index $n=4$ and impact velocities $u=0.2634594,0.2630744,0.2630701$. Note that we identify the bounces by inspecting the scalar field at the origin. We conjecture that a similar resonance mechanism takes place but involving the energy exchange through distinct vibrational modes. According to Fig. 6, there are several small bounces after a larger one, but it is possible to find more significant bounces depending on the correct fine-tuning of the impact velocity. In Fig. 7, we present a 5-bounce resonance for the collision of compactlike structures with $n=16$ and $u=0.20$.

\begin{figure*}[htb]
	\includegraphics[width=6.5cm,height=5.9cm]{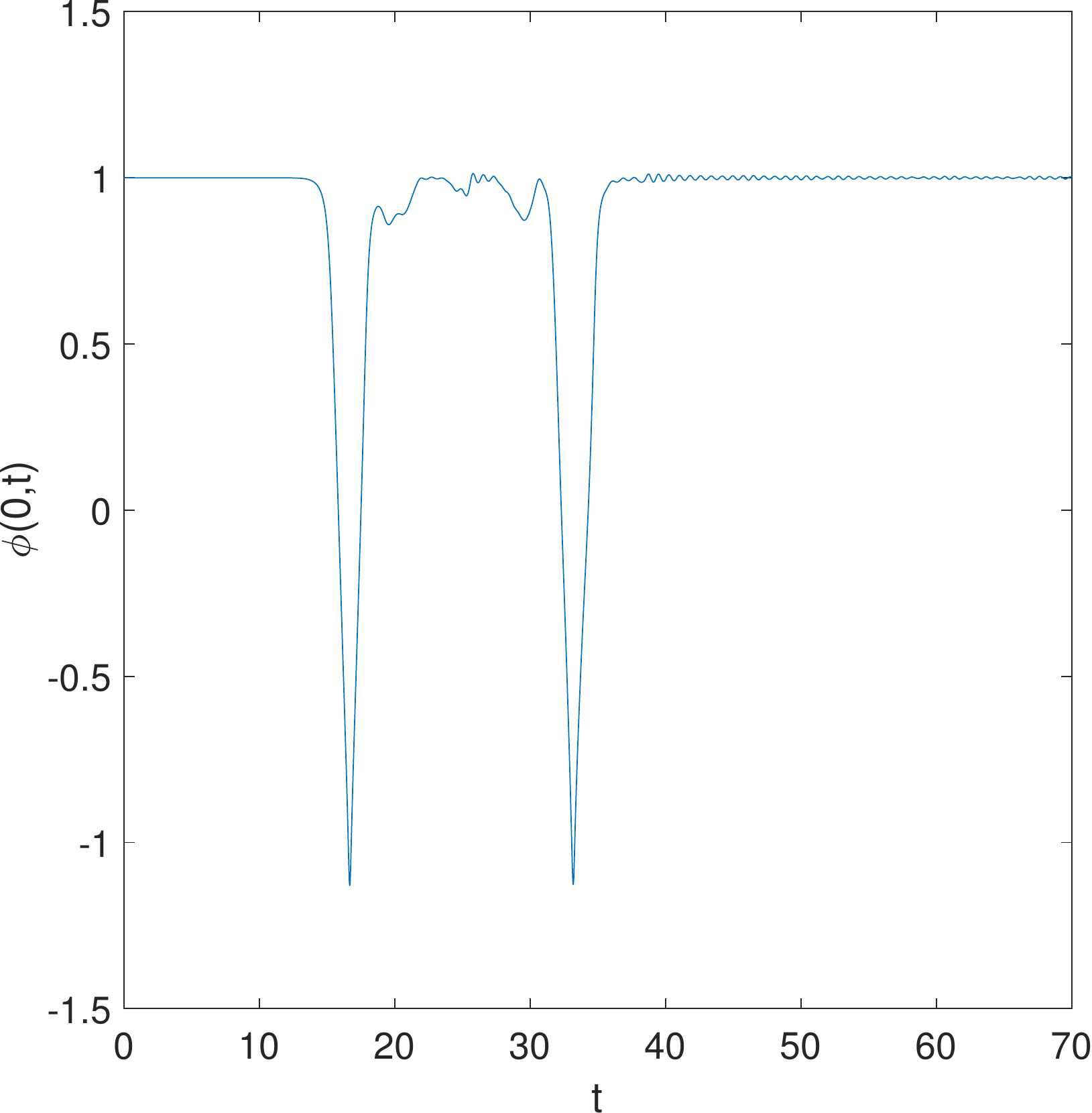}\includegraphics[width=8cm,height=6.5cm]{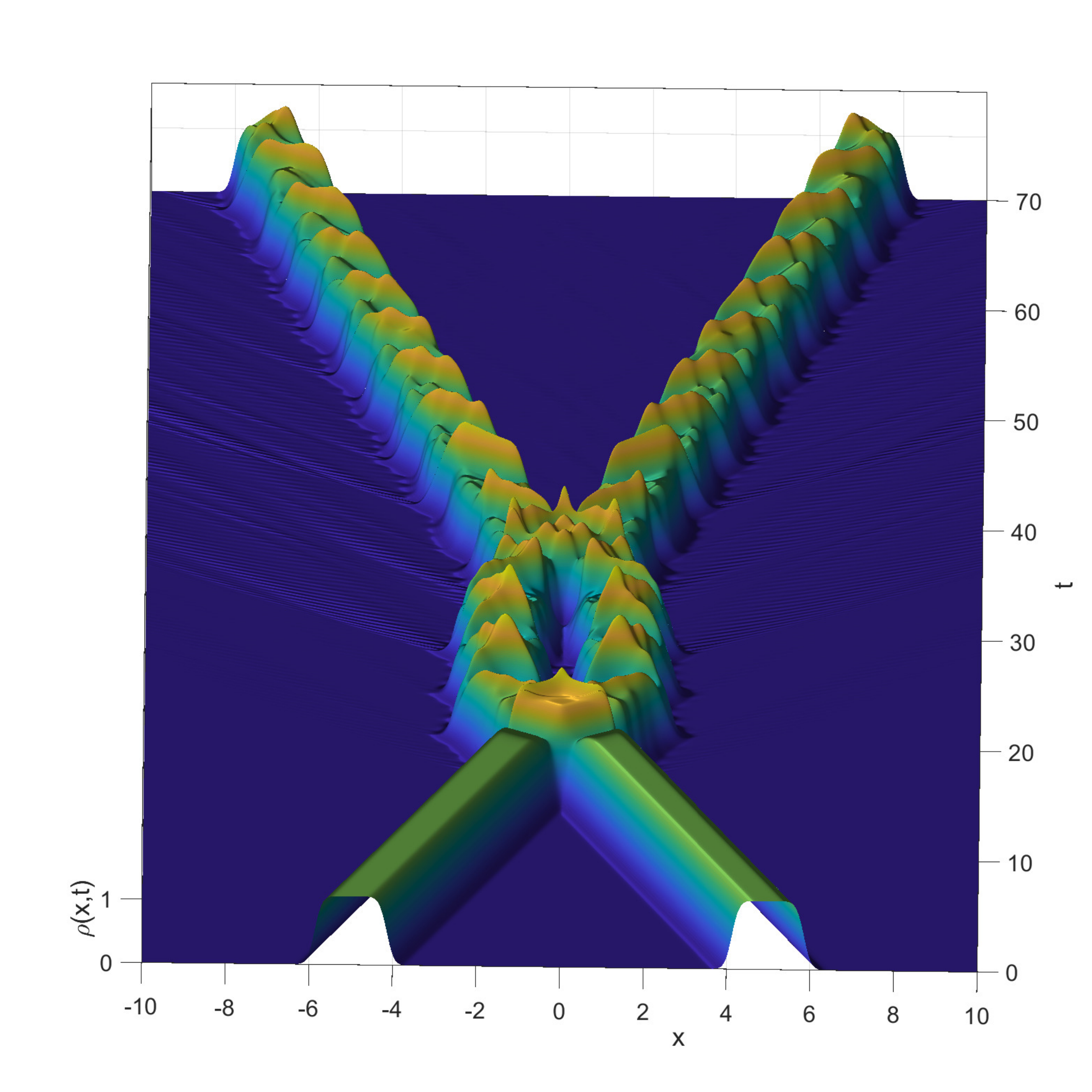}\\
	\includegraphics[width=6.5cm,height=5.9cm]{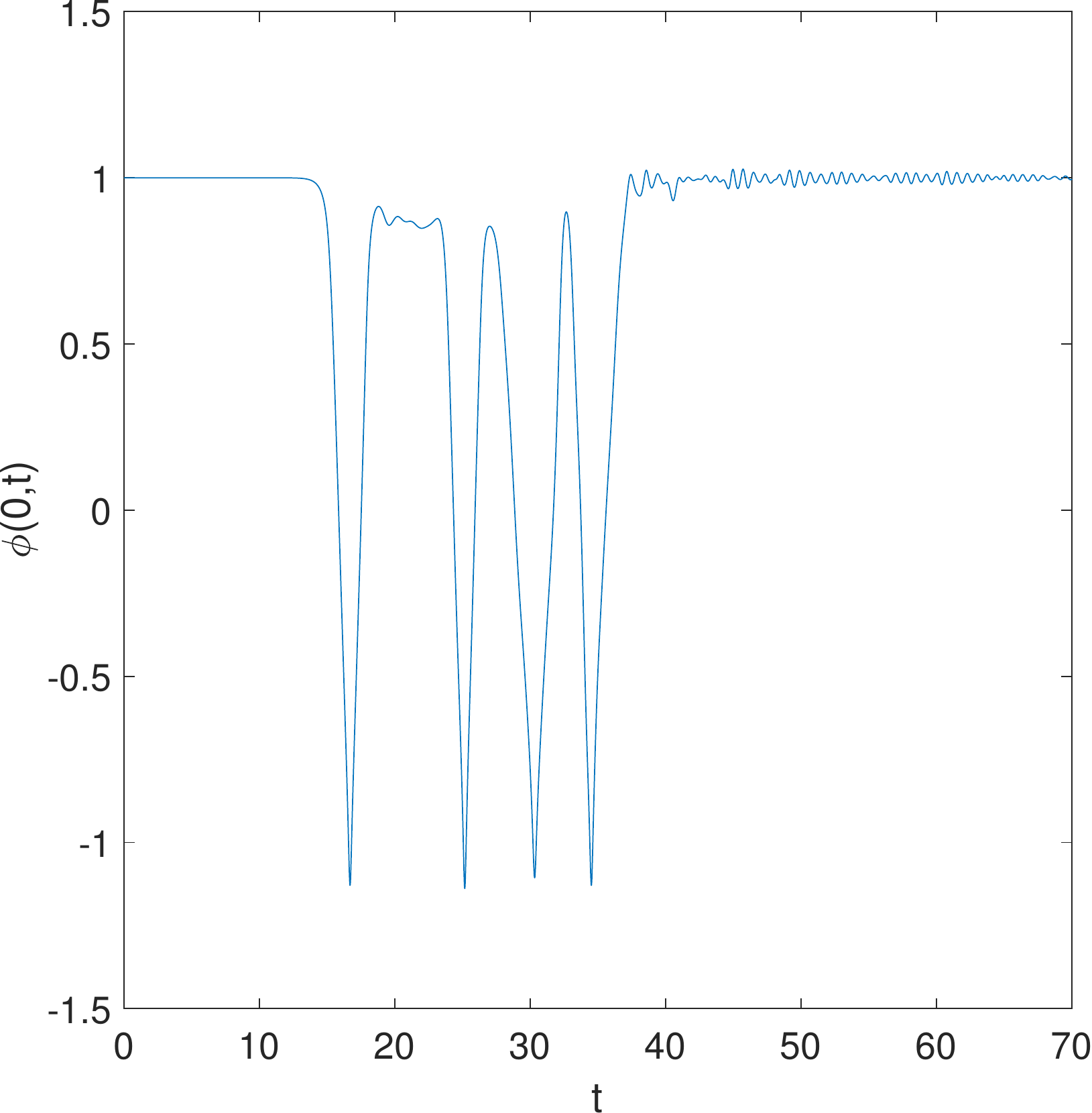}\includegraphics[width=8cm,height=6.5cm]{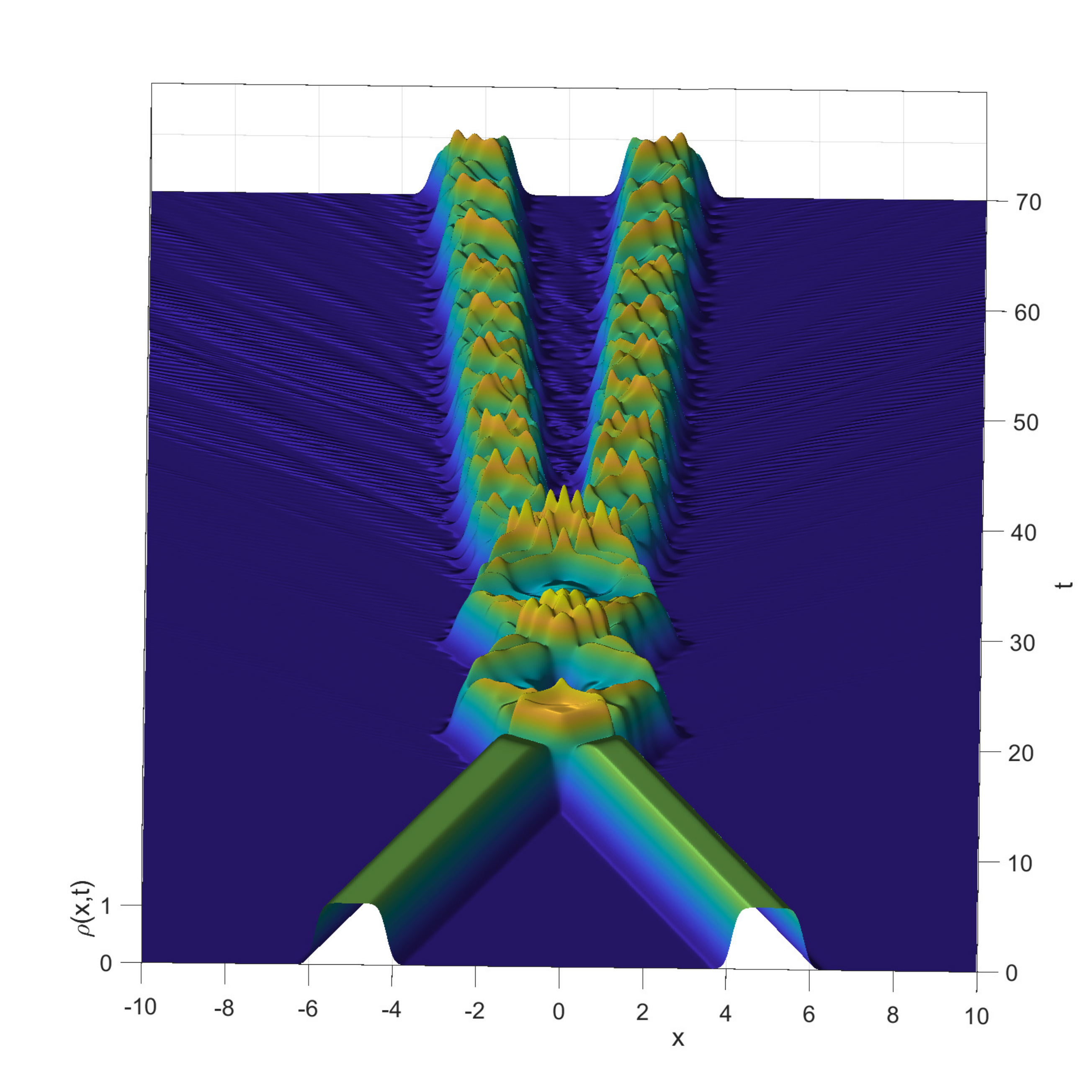}\\
	\includegraphics[width=6.5cm,height=5.9cm]{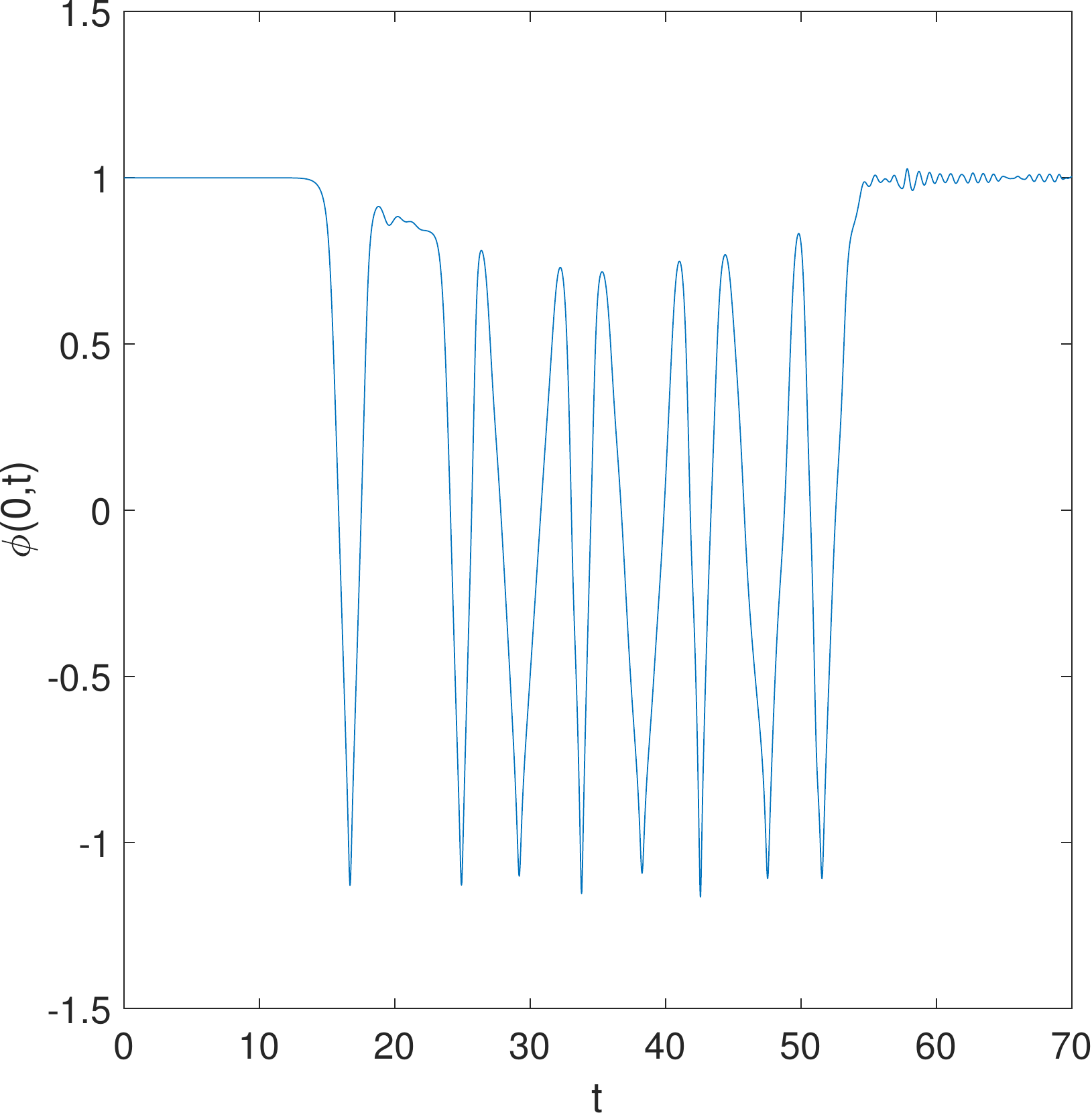}\includegraphics[width=8cm,height=6.5cm]{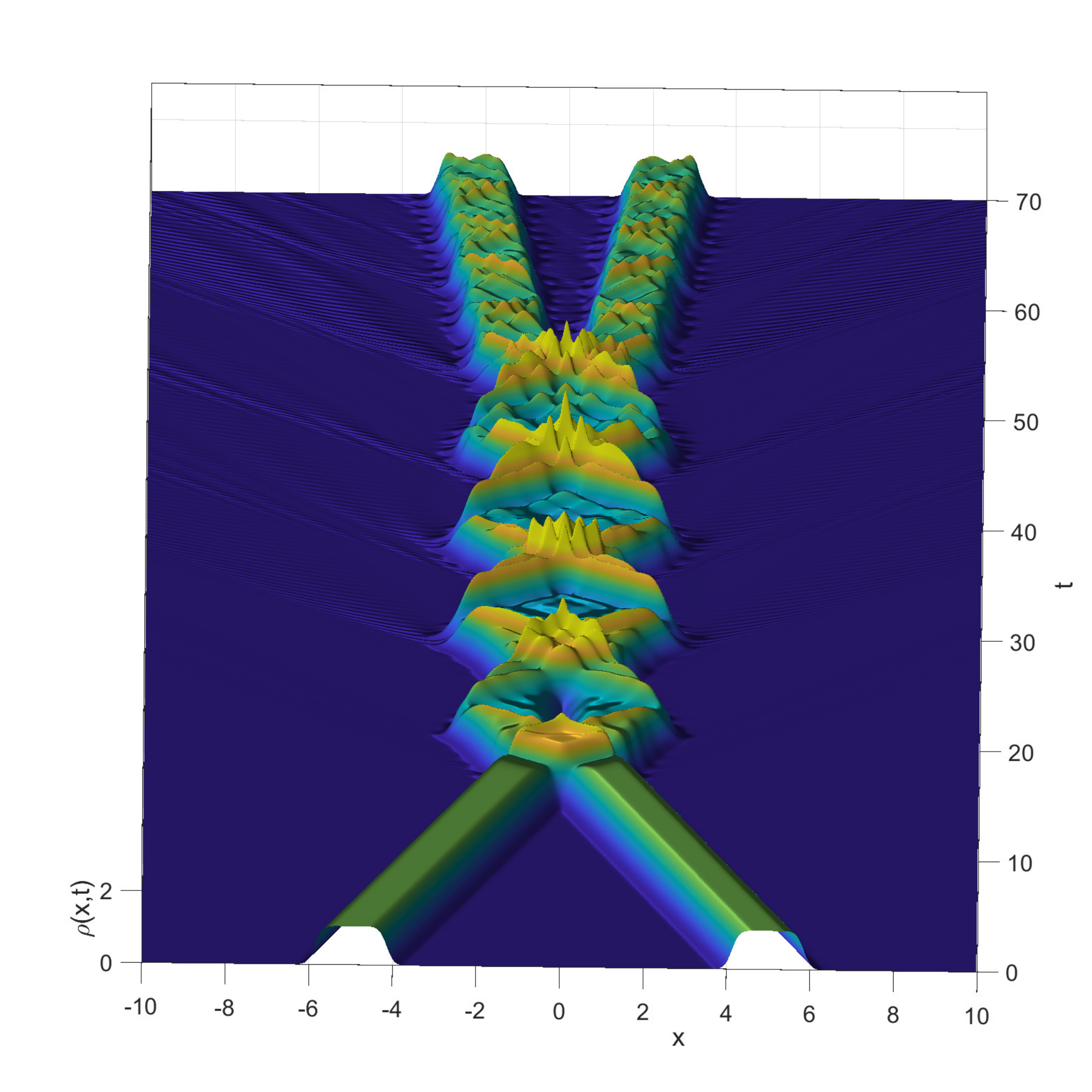}
	\caption{Resonant 2-, 4- and 8-bounce solutions for $n=4$ viewed from the plots of the scalar field at the origin (left panels) and the 3D plots of the energy density (right panels).}
\end{figure*}

\begin{figure}
	\includegraphics[width=6cm,height=5.5cm]{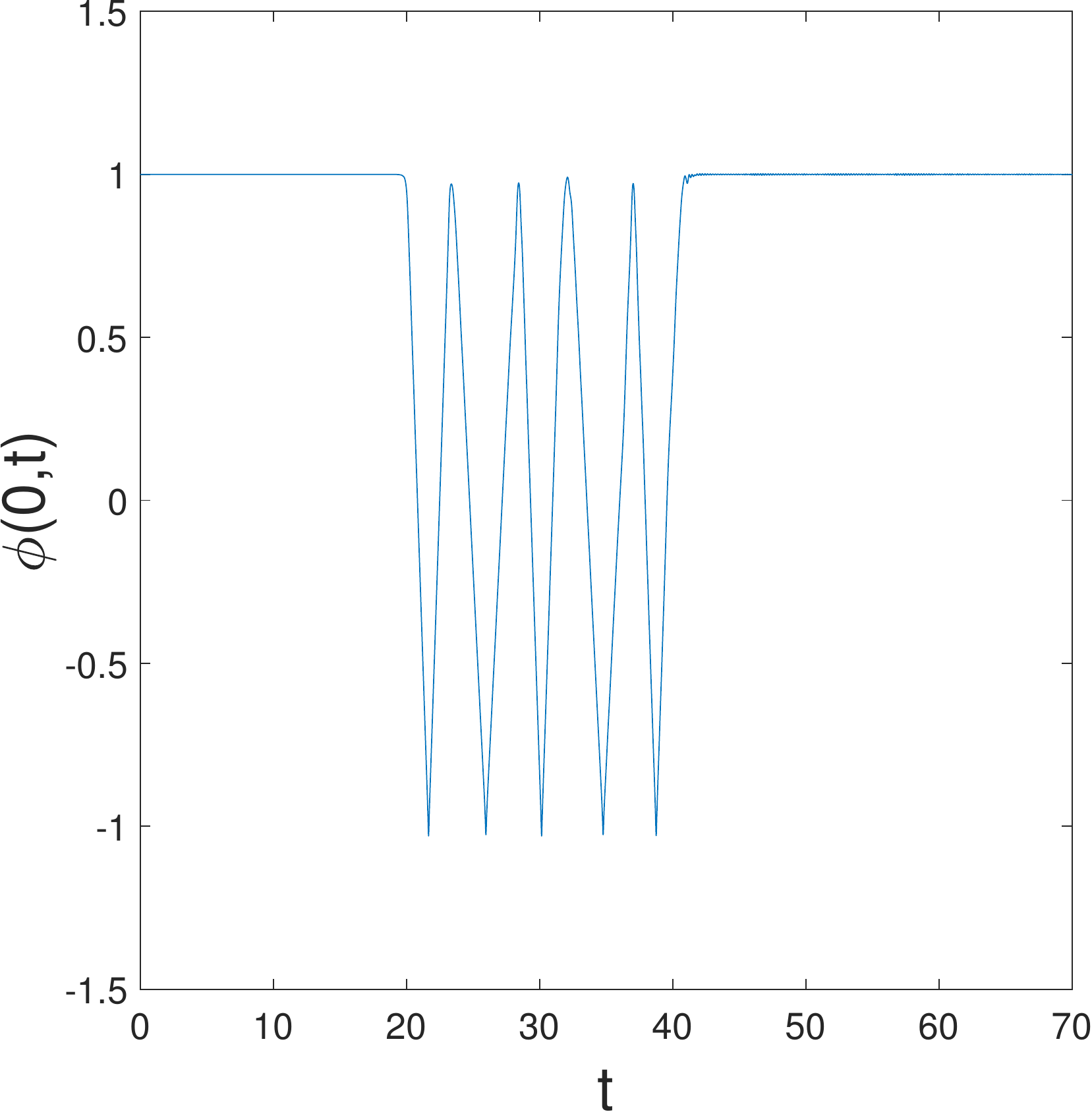}
	\includegraphics[width=8.5cm,height=7cm]{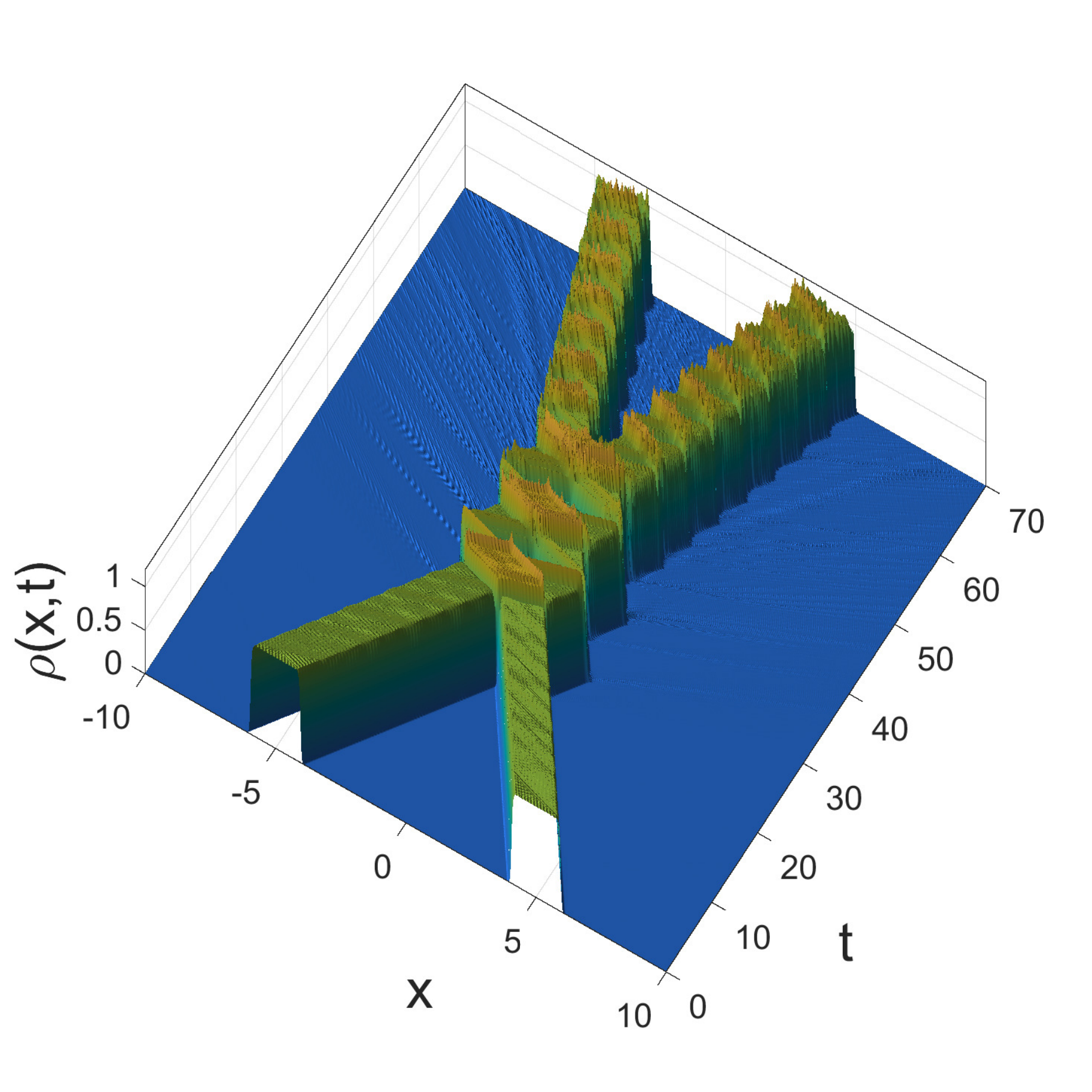}
	\caption{Illustration of a 5-bounce solution for $n=16$.}
\end{figure}

\subsection*{Presence of oscillons}

The collision of kinks is almost elastic, meaning that only a small fraction of the scalar field energy is radiated away. Also, the formation of a bound oscillating scalar field at the origin is the primary outcome for $u < u_{\mathrm{crit}}$. On the other hand, in general, the collision of compactlike defects is inelastic, with a considerable amount of energy of the scalar field radiated. However, there are specific values of the impact velocity where we found the formation of long-lived states we called \textit{compactlike oscillons}. We show a typical compactlike oscillon for $n=4$ and $u=0.08098484$ with the 3d plots of the scalar field and energy density in Fig. 8. 

A new and significant feature we observed after the collision of compactlike kinks is the formation of several structures identified as moving oscillons together with a remnant oscillon at the origin. It is a general outcome for $u < u_{\mathrm{crit}}$ and the pattern of moving oscillons is sensitive to the parameter $n$ ($n > 1$) as well as the impact velocity. Before the formation of the moving oscillons, there is always a period of transient oscillations that can also be interpreted as small bounces. In this case, the energy transferred to the vibrational modes of both compactlike configurations and then back to the translational modes is not enough for the escape of compactlike structures. However, probably due to some resonance mechanism related to the time of unstable oscillations, the transferred energy from the vibrational modes form moving oscillating structures or simply moving oscillons. These oscillons escape in distinct patterns forming a type of cascade. Some examples of the formation of moving oscillons are in Fig. 9 for the collision of compactlike defects characterised by $n=2$, $n=6$ and $n=8$ for $u=0.1816,0.1635$ and $0.25$, respectively.
 
\begin{figure}[htb]
\includegraphics[width=7.2cm,height=4.9cm]{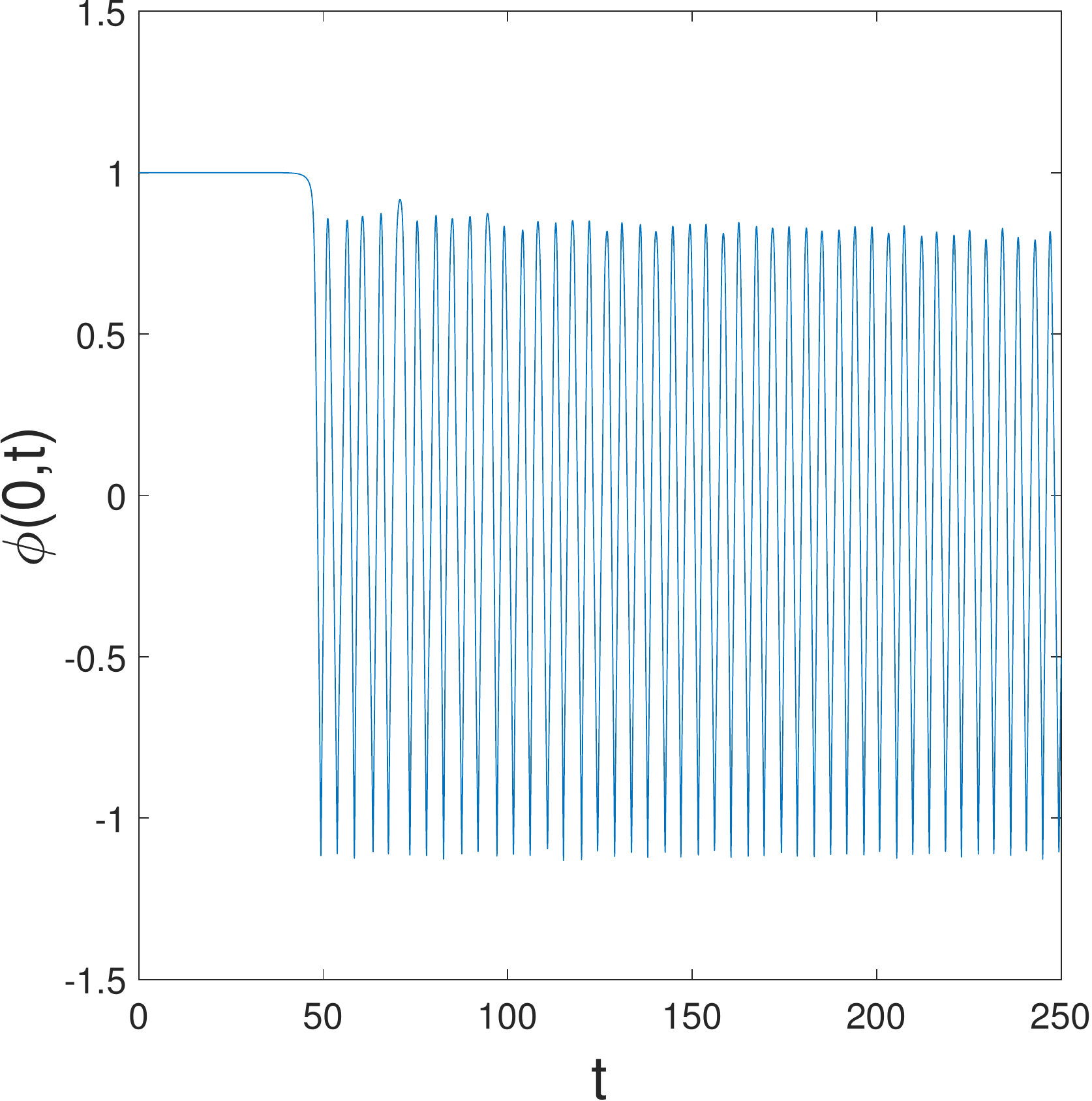}	
\includegraphics[width=9cm,height=7cm]{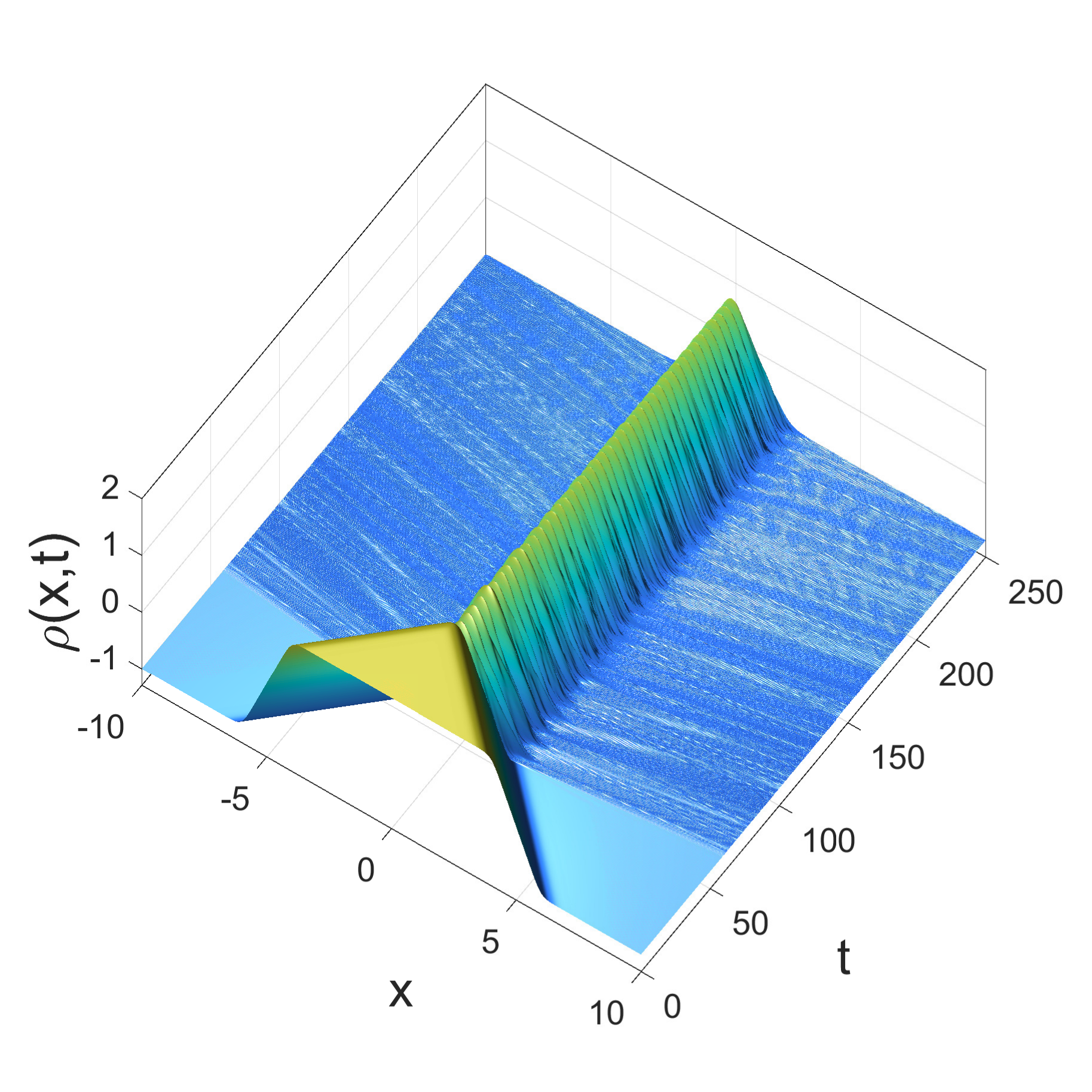}	
\includegraphics[width=9cm,height=7cm]{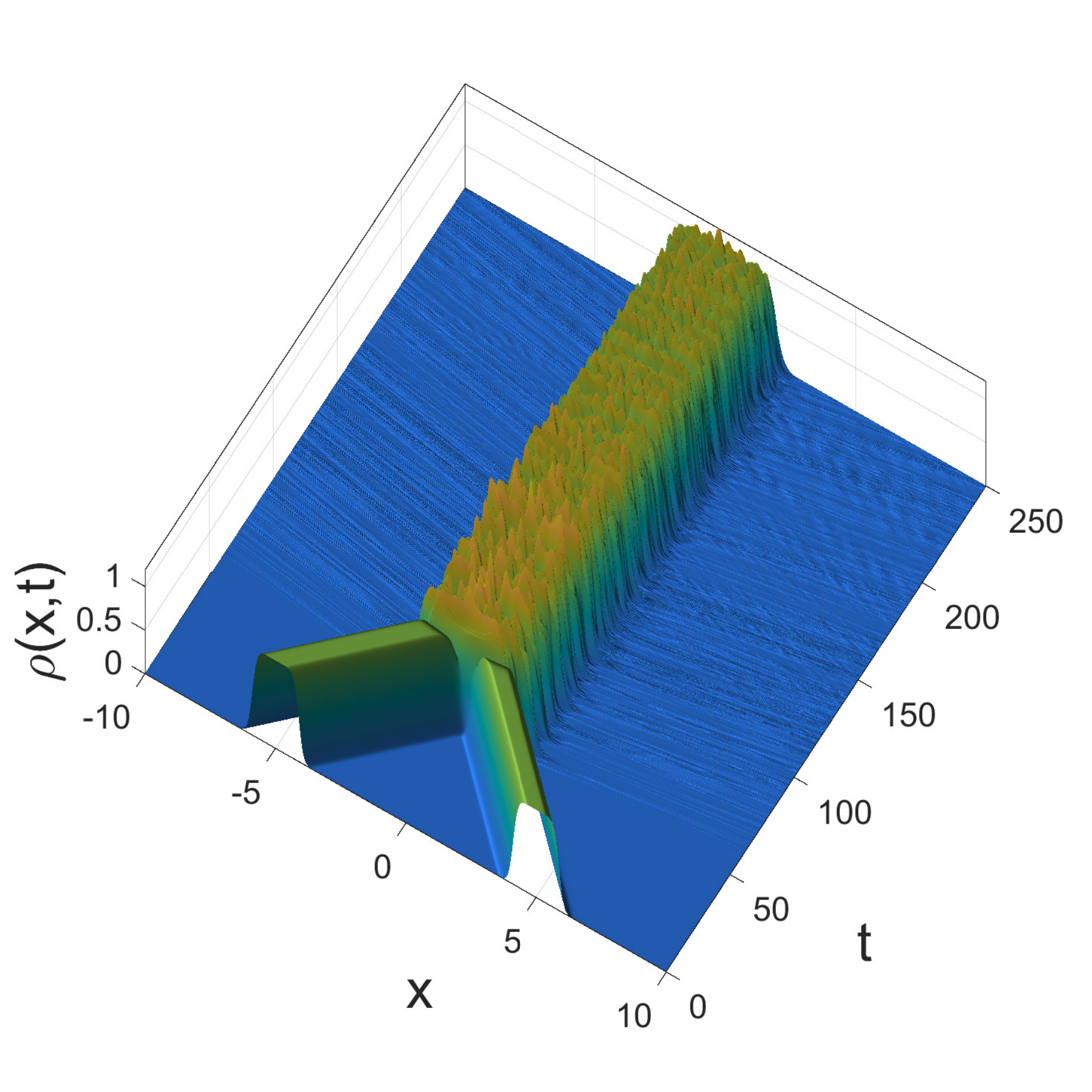}
\caption{Long-lived compactlike oscillon after the collision of compactlike kinks with $n=4$ and $u=0.08098484$.}	
\end{figure}
 
\begin{figure*}[t!]
	\includegraphics[width=7.5cm,height=6.cm]{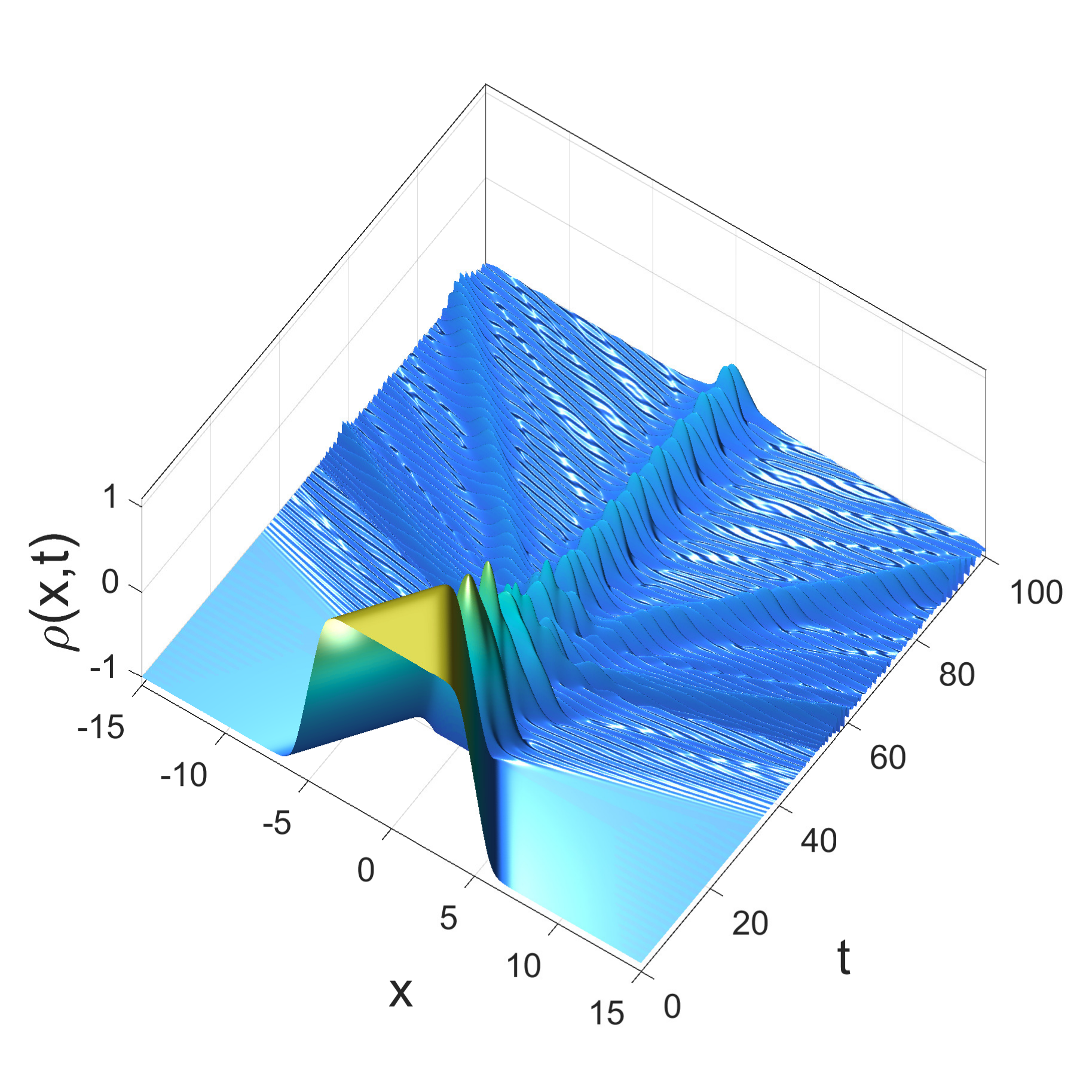}\includegraphics[width=7.5cm,height=6.cm]{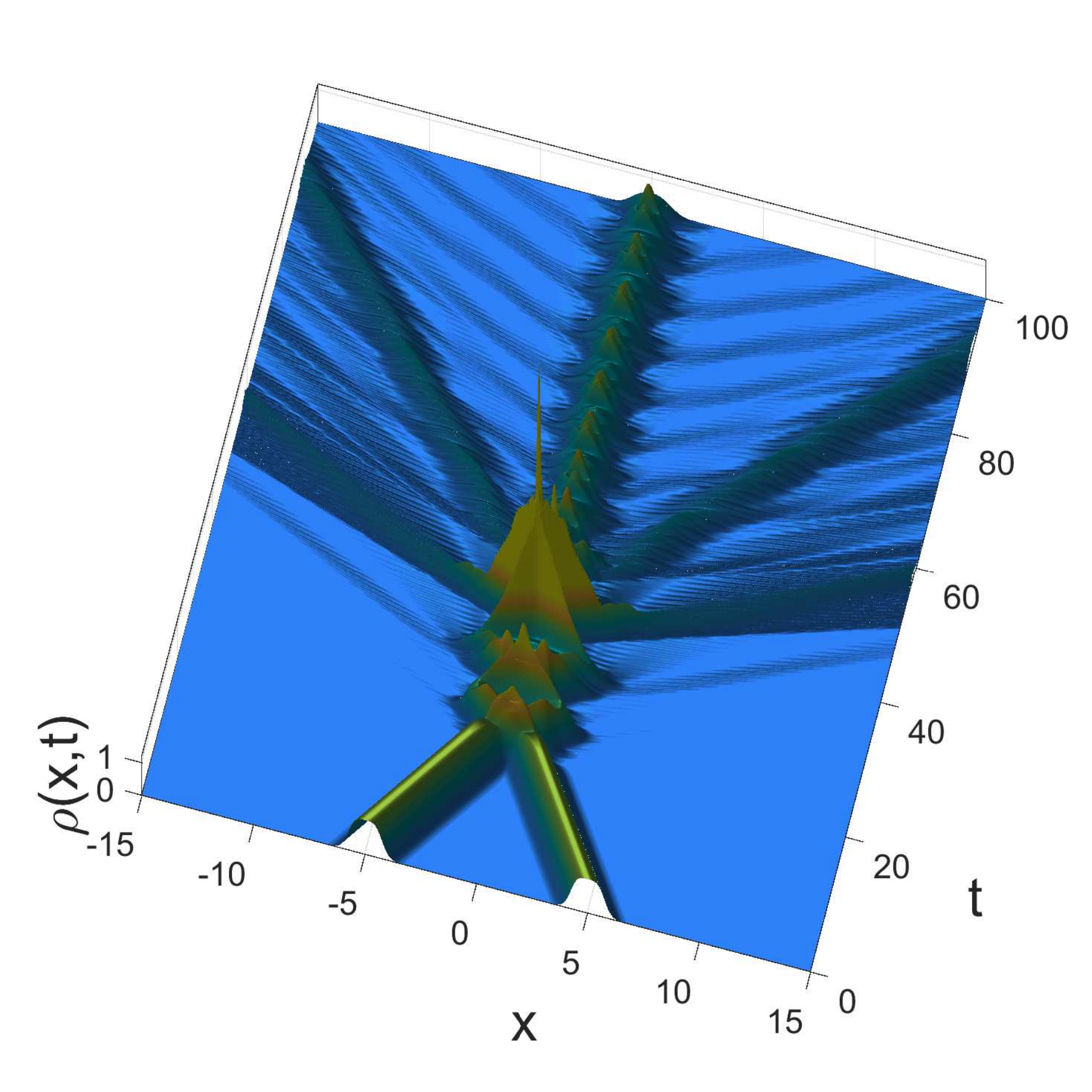}	
	\includegraphics[width=7.5cm,height=7.5cm]{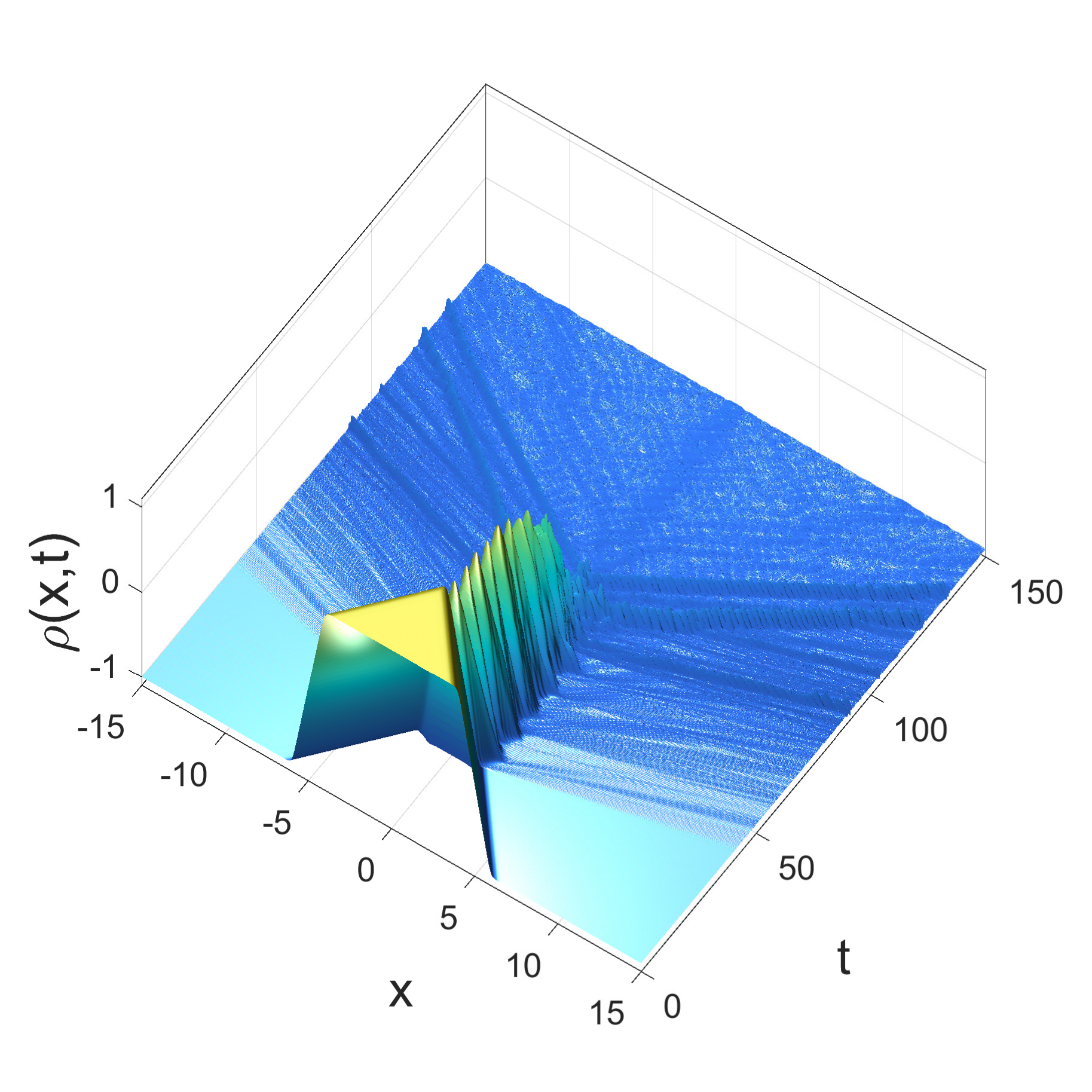}\includegraphics[width=7.5cm,height=6.cm]{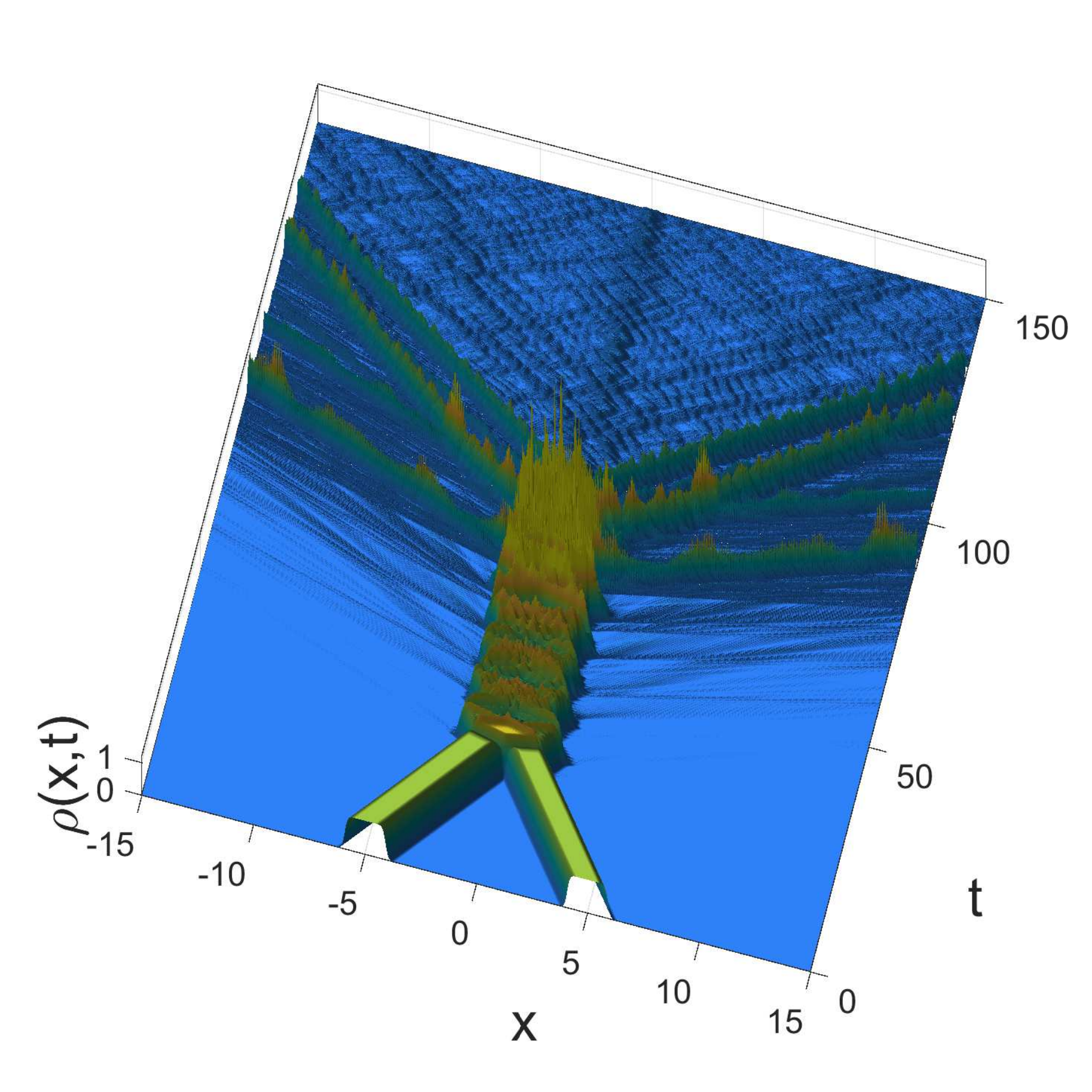}
	\includegraphics[width=7.5cm,height=7.5cm]{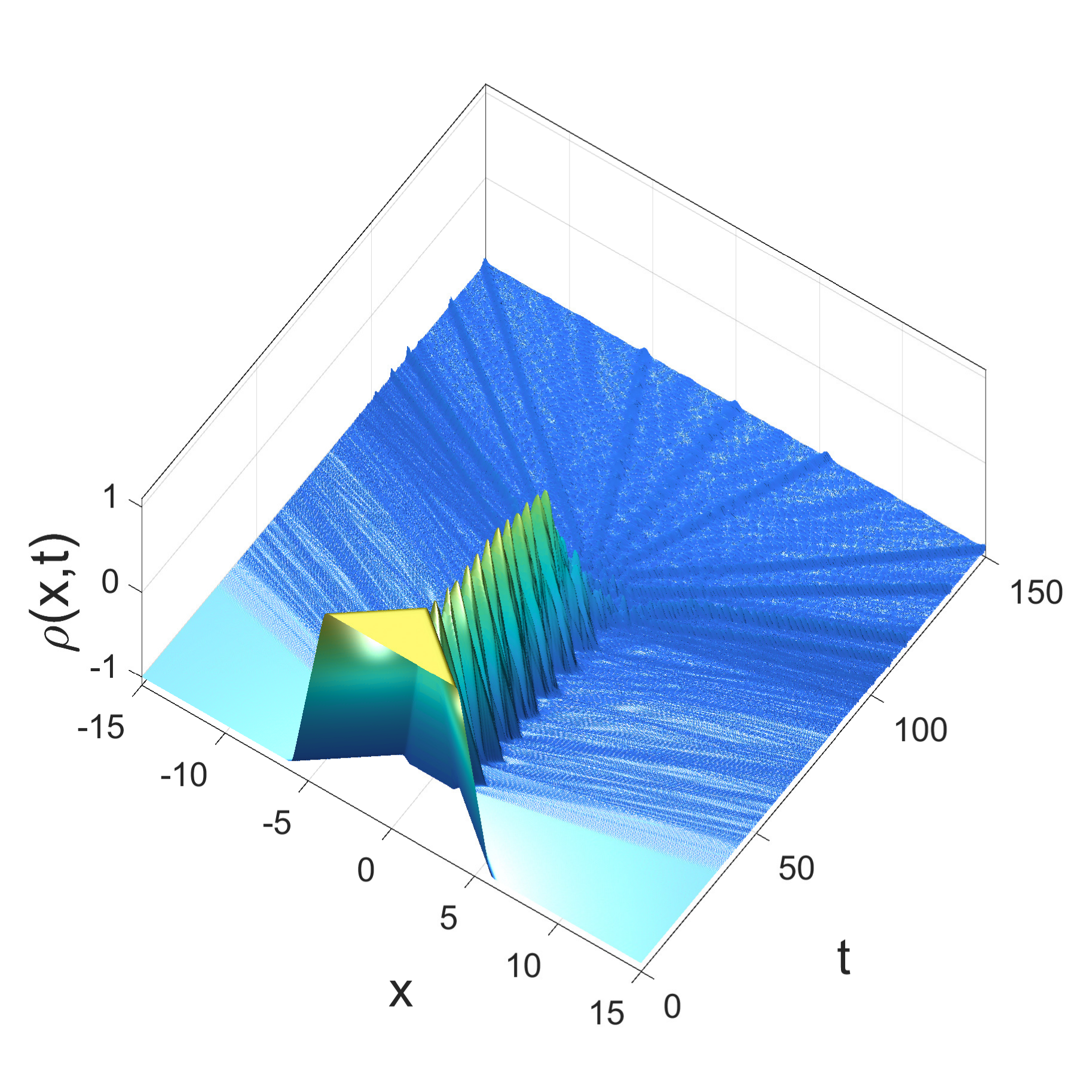}\includegraphics[width=7.5cm,height=6.cm]{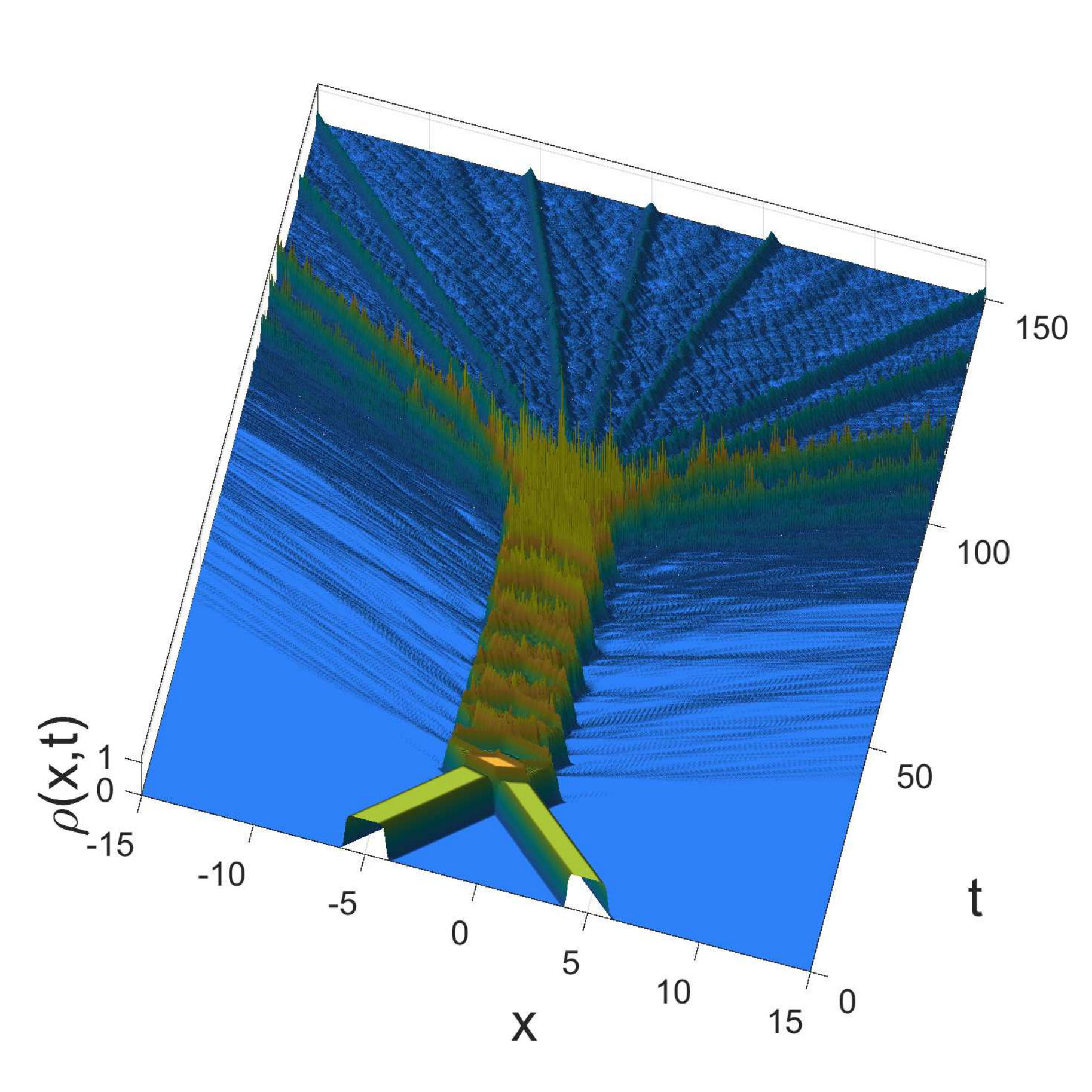}		
	\caption{Top to bottom: moving oscillons generated for the cases $n=2,6$ and $8$.}	
\end{figure*}

\subsection*{Stationary compactlike defects}

The last relevant outcome resulting from fine-tuning the impact velocity at the edges of the bounce windows is the formation of metastable structures, as illustrated in Fig. 10 for $n=8$ and $u=0.2694003$. The compactlike defects are not disrupted or undergo bounces, but emerge and remain to an approximately fixed distance from each other. The time this structure survives depends upon the fine-tuning of the impact velocity. Eventually, the compactlike defects collide again. The metastable configurations can be understood as resulting from the competition between the kinetic energy acquired after the collision that makes the compactlike defects to move apart and the attraction between both defects.  Alternatively, we may interpret the metastable configuration as a consequence of a very long bounce, since eventually, the compactlike defects collide again.  

We remark that the above metastable configuration is not observed for the interaction of kinks ($n=1$). Also, in Ref. \cite{tiago2} the authors reported on the appearance of this type of configuration in the collision of a pair of two-kinks \cite{bazeia}, but there it was identified as a metastable lumplike configuration. This issue requires further investigation.

\section{Conclusion}\label{sec-com}

In this work, we investigated the scattering of a new class of topological structures identified as compactlike defects \cite{b}. They represent a generalization of the $\phi^4$ model with the addition of a parameter $n$ to control  the potential of the model. We recover the kinks of the $\phi^4$ model for $n=1$, whereas for larger values of $n$ the resulting topological defects tend to have a compact profile.

A valuable piece of information for investigating the collision of compactlike defects is the identification of their internal modes. Contrary to the case of kinks, other discrete internal modes appear, with the number increasing as one increases the parameter $n$; see Table 1. This means that during the interaction, energy can be stored and transferred among the internal modes in an intricate manner resulting in a complex dynamics and a richer collision outcome.

\begin{figure}[t!]
	\includegraphics[width=9cm,height=8.cm]{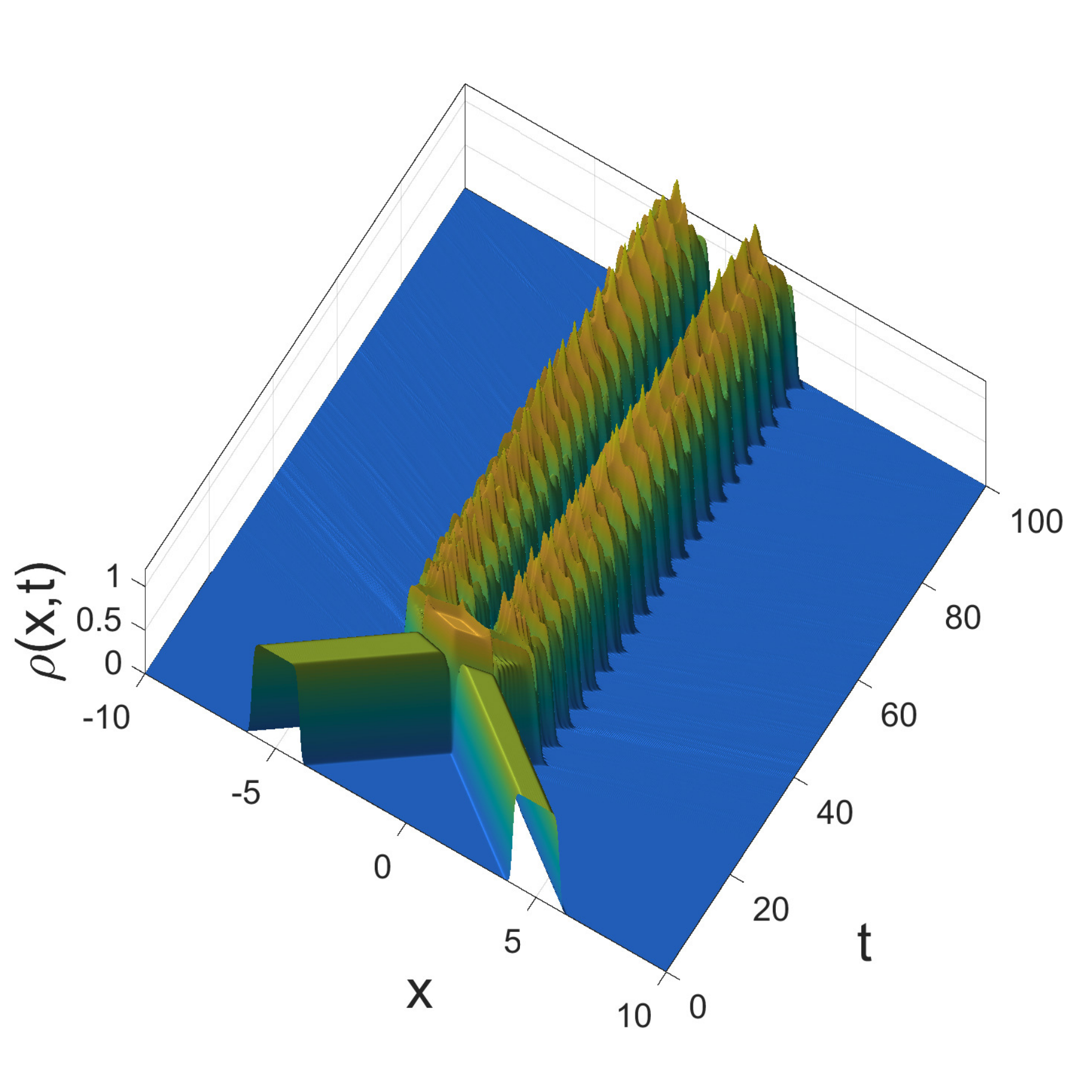}
	\caption{Metastable structure formed by a pair compact-like defects viewed from the 3D plots of the energy density. }
\end{figure}

We described the main results of the compactlike scattering where the choice of the parameter $n>1$ and the impact velocity are the two main factors that dictate the outcome. For any value of $n$, there is a critical velocity, $u_{\mathrm{crit}}$ for which whenever $u>u_{\mathrm{crit}}$, the compact-like defects reflect and move apart from each other. On the other hand, if the impact velocity is smaller than the critical value we have: 

\begin{itemize}
	
\item The compactlike defects can be trapped at the center forming a long-lived oscillating structure we called \textit{compactlike oscillon}. In this case, a small amount of radiation is released. It is similar to the oscillating bound state found in the interaction of kinks.

\item  The prevailing outcome for the impact velocities smaller than the critical velocity is the formation of moving oscillons. Depending on $n$, distinct patterns of moving oscillons are developed and become more intricate as we increase the parameter $n$. We conjectured that a resonance mechanism associated with the several vibrational internal modes is the most probable explanation for the creation of the moving oscillons. Almost all scalar field is radiated away, leaving behind a tiny oscillon at the center.

\item We found a fractal structure at the edges of the windows of escape. It means that even for $u>u_{\mathrm{crit}}$, the compactlike defects start to move apart but collide again to escape in the case of the two-bounce windows. There are also the three-bounce windows, the four-bounce windows, and so on.

\item For those impact velocities close to the edges of the resonance windows, we found the formation of metastable structures in which the compactlike kinks neither escape nor collide again. 

\end{itemize}

The interaction of the more general topological defects displays new features along with the one already known from the case of the kinks of the $\phi^4$ model. We expect that these main outcomes are robust since they are present even for the simplest compactlike defect, characterised by $n=2$. And we also think this would also be the case for the interaction of compactons, in the case of $n\rightarrow \infty$. 

There are several  possibilities of future studies for the present research, and here we point out a direction which is related to the study of the dynamics of two-dimensional defects where the scalar field $\phi=\phi(t,r)$, $r$ being the usual radial coordinate, with the same potential given by Eq. \eqref{potmodel2}. The case $n=1$ has been studied in the literature \cite{gleiser,amin,forgacs,salmi} in connection with the formation of oscillons and application in Cosmology, and we want to investigate how higher values of $n$ contribute to modify the physical picture.

\section*{Acknowledgments}
T. S. Mendon\c ca acknowledges the financial support of the Brazilian agency Coordena\c c\~ao de Aperfei\c coamento de Pessoal de N\'ivel Superior (CAPES). H. P. de Oliveira thanks the Conselho Nacional de Desenvolvimento Cient\'ifico e Tecnol\'ogico (CNPq) and Funda\c c\~ao Carlos Chagas Filho de Amparo \`a Pesquisa do Estado do Rio de Janeiro (FAPERJ) (Grant No. E-26/202.998/518 2016 Bolsas de Bancada de Projetos (BBP)). D. Bazeia and R. Menezes acknowledge CNPq and Paraiba State Research Foundation (Grants 0015/2019 and 0003/2019) for partial financial support.


\begin{thebibliography}{99}
\bibitem{r1} R. Rajaraman, {\it Solitons and instantons}, North-Holland, 1982.

\bibitem{VS} A. Vilenkin and E. P. S. Shellard, {\it Cosmic strings and others topological defects}, Cambridge University Press, 1994.

\bibitem{MS} N. Manton and P. Sutcliffe, {\it Topological solitons}, Cambridge University Press, 2004.

\bibitem{Vacha} T. Vachaspati, \textit{Kinks and domain walls: an introduction to classical and quantum solitons}, Cambridge University Press, 2006.

\bibitem{Shnir}Y. M. Shnir, \textit{Topological and non-topological solitons in scalar field theories}, Cambridge University Press, 2018

\bibitem{AV} A. Vanhaverbeke, A. Bischof and R. Allenspach, Phys. Rev. Lett. { 101}, 107202 (2008).

\bb{DM} G. Basar and G.V. Dunne, Phys. Rev. Lett. {100}, 200404 (2008).

\bb{DS} S. Dutta, D. A. Steer, and T. Vachaspati, Phys. Rev. Lett. {101}, 121601 (2008).

\bb{AI} A. Alonso-Izquierdo, M.A. Gonzalez Leon, and J. Mateos Guilarte, Phys. Rev. Lett. {101}, 131602 (2008).

\bb{TR} T. Romanczukiewicz and Ya. Shnir, Phys. Rev. Lett. {105}, 081601 (2010).

\bb{DMR} P. Dorey, K. Mersh, T. Romanczukiewicz, and Ya. Shnir, Phys. Rev. Lett. {107}, 091602 (2011).

\bb{RH} P. Rosenau and J.M. Hyman, Phys. Rev. Lett. {70}, 564 (1993).

\bb{b} D. Bazeia, L. Losano, M.A. Marques, and R. Menezes, Phys. Lett. B {736}, 515 (2014).

\bibitem{campbell} D. K. Campbell, J. F. Schonfeld and C. A. Vingate, Physica D {9}, 1 (1983).

\bibitem{sca1} V. A. Gani, A. E. Kudryavtsev and M. A. Lizunova, Phys. Rev. D 89  (2014) 125009.

\bibitem{sca2} A. R. Gomes, R. Menezes, K. Z. Nobrega and F. C. Simas, Phys. Rev. D 90 (2014) 065022.

\bibitem{tiago} T. S. Mendon\c ca and H. P. de Oliveira, JHEP 1506 (2015) 133.

\bibitem{sca4} V. A. Gani, V. Lensky and  M. A. Lizunova, JHEP 1508 (2015) 147.

\bibitem{sca5} F. C. Simas, A. R. Gomes, K. Z. Nobrega and J. C. R. E. Oliveira, JHEP 1609 (2016) 104.

\bibitem{sca6} T. Roma\'nczukiewicz, Phys. Lett. B 773 (2017) 295.

\bibitem{sca7}D. Bazeia, E Belendryasova and V. A. Gani, Eur. Phys. J. C 78 (2018) 340.

\bibitem{sca8} E. Belendryasova and V. A. Gani, Commun. Nonlinear Sci. Numer. Simulat. 67 (2019) 414. 

\bibitem{bagomes}D. Bazeia, Adalto R. Gomes, K. Z. Nobrega, and Fabiano C. Simas, Phys. Lett. B 793, 26 (2019).

\bibitem{makhankov} V. G. Makhankov, Phys. Rep. {35}, 1 (1978).
	
\bibitem{moshir} M. Moshir, Nucl. phys. B {185}, 318 (1981).

\bibitem{sugiyama} T. Sugiyama, Prof. Theor. Phys. {61}, 1550 (1979).

\bibitem{aninos} P. Aninos, S. Oliveira and R. A. Matzner, Phys. Rev. D {44}, 1147 (1991).

\bibitem{goodman} R. H. Goodman and R. Haberman, SIAM J. Appl. Dyn. Syst. {4(4)}, 1195 (2005).

\bibitem{tiago2} T. S. Mendon\c ca and H. P. de Oliveira, Braz J Phys (2019). https://doi.org/10.1007/s13538-019-00703-3.

\bibitem{bazeia} D. Bazeia, J. Menezes, R. Menezes, Phys. Rev. Lett. {24}, 241601 (2003).

\bibitem{gleiser} E. J. Copeland, M. Gleiser and H. R. Muller, Phys. Rev. D 52, 1920 (1995); M. Gleiser and D. Sicilia, Phys. Rev. D 80, 125037 (2009).

\bibitem{amin} M. A. Amin, D. Shirokoff, Phys. Rev. D81, 085045 (2010).

\bibitem{forgacs} G. Fodor, P. Forgacs, P. Grandclement and I. Racz, Phys. Rev. D 74, 124003 (2006).

\bibitem{salmi} P. Salmi and M. Hindmarsh, Phys. Rev. D 85, 085033 (2012). 

\end{thebibliography}
\end{document}